\documentclass[12pt]{article} 

\usepackage{hyperref}
\usepackage{url}
\usepackage{setspace}
\usepackage{amsmath,amssymb,amscd,braket}
\usepackage{amsfonts}
\usepackage[x11names]{xcolor}
\usepackage{graphicx}
\usepackage{xcolor}
\usepackage{physics}
\usepackage{bbold}
\usepackage{ytableau}
\usepackage{cancel}
\usepackage{mathtools}
\usepackage{sectsty}
\usepackage{cite}
\allsectionsfont{\boldmath}
\usepackage{tikz}
\def\checkmark{\tikz\fill[scale=0.4](0,.35) -- (.25,0) -- (1,.7) -- (.25,.15) -- cycle;} 

\usepackage{multirow}
\newtheorem{theorem}{Conjecture}

\usepackage{tablefootnote}
\usepackage{footnote}
\makesavenoteenv{table}

 \newcommand{\bea}{\begin{eqnarray}}
\newcommand{\eea}{\end{eqnarray}}
\newcommand{\be}{\begin{equation}}
\newcommand{\ee}{\end{equation}}
\newcommand{\ba}{\begin{align}}
\newcommand{\ea}{\end{align}}

\newlength{\slength}
\newcommand{\ora}[1]{{\color{orange}{ #1}}}
\newcommand{\ma}[1]{{\color{magenta}{ #1}}}
\newcommand{\bl}[1]{{\color{blue}{ #1}}}

\renewcommand{\title}[1]{\vbox{\center\LARGE{#1}}\vspace{5mm}}
\renewcommand{\author}[1]{\vbox{\center#1}\vspace{5mm}}
\newcommand{\address}[1]{\vbox{\center\footnotesize\em#1}}
\newcommand{\email}[1]{\vbox{\center\footnotesize\tt#1}\vspace{5mm}}

\numberwithin{equation}{section}

\newcommand{\mbf}[1]{\mathbf{#1}}
\newcommand{\psiA}{\psi^{A}}
\newcommand{\psiB}{\psi^{B}}
\newcommand{\chiT}[1]{\chi_{\mbf{#1}}}
\newcommand{\chiSU}[1]{\chi^{SU(4)}_{\overline{\mbf{#1}}}}
\newcommand{\SUSU}[2]{(\mbf{#1},\mbf{#2})}

\begin{document}

\begin{titlepage}

\begin{center}

\hfill \\
\hfill \\
\vskip 1cm

\title{Fortuity in ABJM}

\author{Alexandre Belin, Palash Singh, Rita Vadal\`a, Alberto Zaffaroni
}

\address{
Dipartimento di Fisica, Universit\`a di Milano - Bicocca \\
I-20126 Milano, Italy

\vspace{1em}
INFN, sezione di Milano-Bicocca, I-20126 Milano, Italy
}

\email{alexandre.belin@unimib.it, palash.singh@unimib.it, ritalucia.vadala@unimib.it, alberto.zaffaroni@mib.infn.it}

\end{center}

\abstract{

We study $1/12$-BPS and $1/16$-BPS cohomologies and the fortuitous mechanism in ABJM theory. We first establish the existence of fortuitous states in the $N=1$ theory, where the theory is abelian and trace relations are extreme. We then provide explicit constructions of fortuitous states at $N=2$. We find fortuitous states both at weak coupling, in direct parallel to what has been done in $\mathcal{N}=4$ SYM, but we also find additional fortuitous states at $k=2$, which is in the strongly coupled regime. The extra fortuitous states that appear at $k=2$ are in non-trivial monopole sectors. A striking distinction from $\mathcal{N}=4$ SYM is that the fortuitous states appear at much smaller quantum numbers, making them easier to find. Along the way, we formulate a non-renormalization conjecture for cohomologies in ABJM.

}

\vfill

\end{titlepage}

\eject

\tableofcontents

\section{Introduction}

Black holes provide a theoretical laboratory to explore the inner workings of quantum gravity. Bekenstein and Hawking postulated that black holes carry an entropy proportional to the area of their event horizon \cite{Bekenstein1,Hawking1}
\be
S_{BH}= \frac{A_H}{4G_N} \,,
\ee
which implies that from outside, black holes behave as standard quantum mechanical system with a Hilbert space of dimension $e^{S_{BH}}$. The Bekenstein-Hawking formula counts the number of microstates of quantum gravity. Remarkably, for certain supersymmetric black holes, a microscopic counting of these microstates was achieved. This was first done for black holes in asymptotically flat space, as pioneered by Strominger and Vafa \cite{Strominger:1996sh}. More recently, this was also achieved for black holes in asymptotically AdS space, first in AdS$_4$ \cite{Benini:2015eyy}, and shortly thereafter in AdS$_5$ \cite{Benini:2018ywd,Cabo-Bizet:2018ehj,Choi:2018hmj}. The matching is done between the degeneracies in the superconformal index and the area of the black hole with the corresponding charges, and crucially relies on supersymmetry as the index is invariant under continuous coupling deformations, or alternatively can be computed by localization.

For AdS black holes with spherical horizons, the objects being counted are clear: in the dual CFT, they correspond to BPS local operators with large charges, scaling with the appropriate power of the rank of the gauge group $N$. While we understand well which objects are being counted, the precise nature of these microstates remains largely elusive. Supersymmetry provides hope that one can go beyond a mere counting of the states, and that a more detailed understanding of their structure is within reach.

A program with this aim, for small values of $N$, was initiated in \cite{Chang:2022mjp} building on the original attempt of \cite{Chang:2013fba}. An important point is that one should distinguish different classes of BPS operators: some correspond to (multi)-graviton states, and they should not be mistaken with true black hole microstates. The difference between the two lies on a key point proposed in \cite{Chang:2024zqi}: graviton states have a nice large $N$ limit, and remain BPS at all values of $N$. They are dubbed \textit{monotone} states. On the contrary, true black hole microstates are erratic objects in $N$-space, and a BPS black hole operator at some value of $N$ will fail to stay BPS at some larger value of $N$. They are called \textit{fortuitous}. The erratic nature of these operators, as viewed in $N$ space, resonates with the chaotic nature expected of black holes \cite{Schlenker:2022dyo}. 

At the technical level, the origin of the erratic features in $N$ are the trace relations in the boundary gauge theory. These relations fulfill several roles: they impact the graviton spectrum by truncating the single-trace spectrum while also inducing relations between multi-graviton states, that would have been independent at infinite $N$. This results in a finite $N$ graviton spectrum which is a strict subset of its infinite $N$ counterpart (the bulk Fock space). But importantly, the trace relations do more than this: they also allow for extra BPS operators, which are absent in the infinite $N$ theory. These are the fortuitous black hole operators.

Fortuity has been studied in several theories relevant for black holes in AdS/CFT, as we summarize in Table \ref{table1}: $\mathcal{N}=4$ SYM \cite{Chang:2022mjp,Chang:2024zqi,Choi:2022caq,Choi:2023vdm,Choi:2023znd,Budzik:2023vtr}, the D1-D5 CFT \cite{Chang:2025rqy,Hughes:2025car,Chang:2025wgo}, and $\mathcal{N}=2$ SYK \cite{Chang:2024lxt}. It has also been studied in simpler theories, such as vector models \cite{deMelloKoch:2025cec,Kim:2025vup} or even in a simple fermionic matrix theory in quantum mechanics \cite{Chen:2025sum}.\footnote{The fortuity mechanism is part of a larger program aiming to understand the nature of trace relations and finite $N$ Hilbert space, in particular the giant graviton expansion. See \cite{Gaiotto:2021xce,Murthy:2022ien,Liu:2022olj,Eleftheriou:2023jxr,Lee:2024hef,Lee:2025veh,deMelloKoch:2025rkw,Caputa:2025ikn,deMelloKoch:2025eqt,Hatsuda:2024lcc} for more details. } Within the canonical and maximally supersymmetric  examples of AdS/CFT, the two theories yet to consider are ABJM and the six-dimensional $(2,0)$ theory. The goal of this paper is to fill the gap for ABJM, and study its fortuitous states. ABJM theory has several properties that have no direct counterpart in $\mathcal{N}=4$: there is no continuous coupling, so the supercharge cohomologies are allowed to jump as the coupling is varied. There are also monopole operators present, some of which will necessarily be fortuitous. Finally, the theory has both a type IIA limit and a M-theory limit, which may have different fortuitous properties. 

\begin{table}[!h]
\begin{center}
\label{table1}
\begin{tabular}{|c| c| c | c|} 
 \hline
 d & Theory & Analyzed & Specificity \\
 \hline\hline
 1 & $\mathcal{N}=2$ SYK & \checkmark &  \\ 
 \hline
 2 & D1-D5 & \checkmark & Trace-relation simple   \\
 \hline
 3 & ABJM & {\color{green} \checkmark} & No continuous coupling + monopole operators  \\
 \hline
 4 & $\mathcal{N}=4$ SYM & \checkmark & Benchmark  \\
 \hline
 6 & $(2,0)$ Theory & & Non-Lagrangian + M-theory limit only\\  
 \hline
\end{tabular}
\end{center}
\caption{A summary of the various holographic theories and the status of the fortuity search, along with what makes the theories different from the original search in $\mathcal{N}=4$ SYM. In the D1-D5 CFT, the trace relation is the stringy exclusion principle \cite{Maldacena:1998bw} which is much simpler than trace relations in matrix theories. In ABJM, there is no continuous coupling, there are monopole operators and there is also an M-theory limit. In the $(2,0)$ CFT, there is only an M-theory limit with no coupling constant, and the theory is non-Lagrangian. In this paper, we fill the gap for ABJM theory.}
\end{table}

\subsection{Summary of results}

In this paper, we provide a detailed analysis of fortuitous operators in $U(N)_k \times U(N)_{-k}$ ABJM theory. We analyze the 1/12-BPS coholomologies (or 1/16-BPS at $k=1,2$), working with various values of $N$ and $k$, and the outcome of the analysis depends on the choice of parameters. The $N=1$ case is somehow special because  the action of the supercharge on the BPS letters vanishes, and therefore the cohomological problem is trivial. Nevertheless, adhering to the definition given above, we find no fortuity when $k=1$, while we find a huge number of fortuitous states for $k>1$. We can see explicitly that they disappear as cohomologies at large $N$. At $N=2$ the cohomological problem becomes non-trivial. We find fortuitous states at low order in the expansion parameter $x$ for the index. They can be identified unambiguously for large $k$, where monopoles are absent. At finite $k$, monopole operators appear in the spectrum and can change the number of fortuitous operators. We focus on $k=2$ where we can find monopole operators in the cohomology at low order in $x$ and, with a combinations of arguments, we find the full set of fortuitous states at order $x^3$. We do not consider the case $k=1$ where monopole operators hit the unitary bound and the theory contains a free sector, since we expect a strong similarity with  $\mathcal{N}=4$ SYM where fortuity appears at much higher order. In more details, we find the following results.

\subsubsection*{$N=1$ and $k=1$}

Unlike, $\mathcal{N}=4$ SYM, ABJM theory is a non-trivial CFT at $N$=1, and it can be strongly coupled at small $k$, even if its gauge component is abelian. We show that there are no fortuitous states in ABJM at $N=1$ and $k=1$. This follows from the fact that the index is freely generated, and the full cohomology with these parameters can be embedded as a strict subset of the infinite $N$ cohomology.

\subsubsection*{$N=1$ and $k>1$}

Here, we find a huge number of fortuitous states. The simplicity of the superconformal index in the abelian theory makes it possible to specify very clearly the nature of the fortuitous states. We also find a feature that, to the best of our knowledge, has not appeared in any other study on fortuity: we find single-trace\footnote{It is slightly subtle to define single versus mutli- trace operators in the $N=1$ theory, but here by single-trace we mean that it is not a composite of BPS operators of lower scaling dimension.} fortuitous states. These appear already at second non-trivial order in the superconformal index. The superconformal index has an interesting property: it can fully be accounted by second-quantizing the set of single-trace operators, and keeping track of relations. But some of these single-trace operators are not BPS at higher values of $N$. Both this property of the index and the fortuitous states we find can ultimately be seen as following from the extreme nature of trace relations at $N=1$, even more extreme than those found in the ABJ theory \cite{Kim:2025vup}.

\subsubsection*{$N=2$ and large $k$}

This is the value of the parameters that resemble the most the analysis that has been done in $\mathcal{N}=4$ SYM, representing a weakly coupled limit. We work with the one-loop supercharge, and formulate a non-renormalization conjecture that prevents any lifting beyond one loop, even if the coupling is not continuous. We find perfect agreement with the superconformal index. We then provide a complete classification of the cohomologies up to the first three non-trivial orders in the superconformal index. At the third order, the first trace relations kick in, and it is also at this order that we identify the first fortuitous states.\footnote{It is worthwhile to note a distinction with $\mathcal{N}=4$ SYM here, where the trace relations kick in much earlier than the appearance of fortuitous states.} The fortuitous operators are
\begin{equation}
    \mathcal O_f = \epsilon^{ij}\left(\psiA_iA_jB_kA_l+\psiB_iB_jA_lB_k\right)-\frac{3}{4}\,\epsilon^{ij}\left(B_kA_l\right)\left(\psiB_iB_j+\psiA_jA_i\right)  \,,
    \label{eq:fortuitous_intro}
\end{equation}
for $i,j=1,2$.  We show that these operators are $Q$-closed due to a trace relation valid only at $N=2$. At higher $N$, they would thus not be BPS and are therefore fortuitous. With the inclusion of these four operators, we correctly match the superconformal index. The fortuitous operators are considerably simpler than in $\mathcal{N}=4$ SYM, which follows from the fact that they appear at much lower order in the superconformal index. They share similarities with the lowest fortuitous operator in $\mathcal{N}=4$, which can be written as an operator in the Konishi multiplet, dressed with two gravitons \cite{Budzik:2023vtr}. There, it is triple-trace. Here, we find a double-trace operator: the dressing by the lightest graviton of a non-BPS operator at second order. Our operator also has the new feature of having a single-trace component to it. 

\subsubsection*{$N=2$ and $k=2$}

We also analyze the fortuitous operators at small $k$, and focus on $k=2$ where supersymmetry is enhanced to $\mathcal{N}=8$. We study the cohomologies to third non-trivial order in the superconformal index, and identify ten fortuitous states at that order. Given that understanding relations among monopole operators and finding the  action of the supercharge in non-trivial monopole sectors is difficult, we use the enhancement of the R-symmetry to $SO(8)$ as a guide. The choice of a supercharge  breaks the symmetry preserved by the superconformal index and the cohomology down to $SU(4)$. We then assemble operators into representations of $SU(4)$ and find that the multi-graviton index mismatches the graviton count by 10 fermionic states. These 10 fermionic states are the fortuitous cohomologies and we propose that they are the $SU(4)$ completion of the four fortuitous operators that we found at large $k$.

The rest of the paper is organized as follows. In Section \ref{sec:2}, we review ABJM theory and define the appropriate supercharge cohomology and the appropriate BPS letters. In Section \ref{sec:3}, we study the cohomologies for $N=1$. In Section \ref{sec:4}, we study the cohomologies for $N=2$ and large $k$. In Section \ref{sec:5}, we discuss the case $N=2$ and $k=2$. We conclude with some discussion and open questions in Section \ref{sec:6}. Three appendices contain technical details.

\section{The $U(N)_k \times U(N)_{-k}$ ABJM theory \label{sec:2}}

The Aharony--Bergman--Jafferis--Maldacena (ABJM) theories form a large class of three-dimensional superconformal field theories,  with $\mathcal N=6$ supersymmetry. We will be interested in the $U(N)_k \times U(N)_{-k}$ ABJM theories, realised as Chern--Simons matter theories with gauge group $U(N)\times U(N)$ and Chern-Simons level $k$ and $-k$. The Lagrangian is given in Appendix \ref{app:ABJM}.

In the large $N$, finite $k$, limit, these are holographically dual to M-theory on $AdS_4\times S^7/\mathbb Z_k$. There is also another holographic limit. If instead we work in the 't Hooft limit, i.e. $N,k \rightarrow \infty$, with $\lambda=N/k$ fixed, the dual is described by type IIA string theory on $AdS_4\times \mathbb{CP}^3$. The quantity $\lambda$ acts as the effective coupling constant for the theory, so it is weakly-coupled for $\lambda\ll1$, while strongly coupled for $\lambda\gg 1$ which is the holographic point. 

We start by reviewing some basic properties of the theory, in particular its supersymmetry and quiver description.

\subsection{Quiver and supersymmetry}

The $U(N)_k\times U(N)_{-k}$ ABJM theory is a 3d $\mathcal N=6$ superconformal field theory, and hence admits the $\mathfrak{osp}(6\vert 4)$ superconformal algebra. This has a bosonic subalgebra $SO(3,2)\oplus SO(6)_R$ where the first piece is the 3d global conformal group and $SO(6)_R \cong SU(4)_R$ is the $R$-symnmetry algebra.\footnote{Note that this is only true at the level of the Lie algebra. We do not distinguish between the Lie group and the Lie algebra throughout this paper.} It is well known that for $k=1,2$ there is a non-perturbative supersymmetry enhancement to $\mathcal N=8$ supersymmetry.

As a 3d $\mathcal N=6$ SCFT, ABJM contains 12 Poincare supercharges $\left(Q_{IJ}\right)_\alpha$ and the corresponding 12 conformal supercharges $\left(S^{IJ}\right)^\alpha$, where $I,J\in\{1,2,3,4\}$ are antisymmetric indices and $\alpha\in\{\pm\}$. These satisfy the relations
\begin{equation}\label{eq:algebra}
    \begin{split}
        \left\{\left(Q_{IJ}\right)_\alpha\,,\,\left(Q_{KL}\right)_\beta\right\} = \epsilon_{IJKL} P_{\alpha\beta} \;\;&, \quad \left\{\left(S^{IJ}\right)^\alpha\,,\,\left(S^{KL}\right)^\beta\right\} = \epsilon^{IJKL} K^{\alpha\beta} ~, \\[1.5mm]
        \left\{\left(Q_{IJ}\right)_\alpha\,,\,\left(S^{KL}\right)^\beta\right\} = -4\,\delta_\alpha^\beta \delta_{[I}^{[K} R_{J]}^{L]}& -2\,\delta_{[J}^K\delta_{I]}^L \, M_\alpha^\beta - 2\,\delta_{[J}^K\delta_{I]}^L \delta_\alpha^\beta \, \mathcal D ~,
    \end{split}
\end{equation}
where $M_\alpha^\beta$, $P_{\alpha\beta}$, $K_{\alpha\beta}$, $\mathcal D$ are the Lorentz, translation, special conformal generators, and Dilation generators respectively, while $R_I^J$ are the $SU(4)_R$ $R$-symmetry generators. In radial quantisation, the Poincar\'e and conformal supercharges are related by hermitian conjugation, $\left(Q_{IJ}\right)_\alpha^\dagger = \left(S^{IJ}\right)^\alpha$.

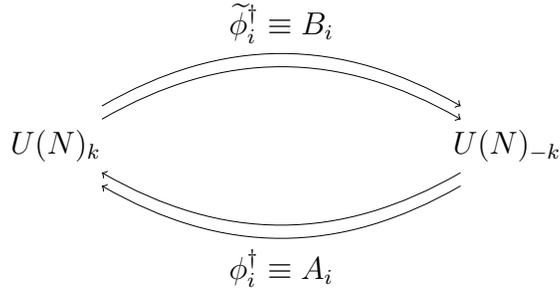
\begin{figure}[!h]
    \begin{center}
        \begin{tikzpicture}
            \node (L) at (-3,0) {$U(N)_k$};
            \node (R) at (3,0) {$U(N)_{-k}$};
	        \draw[line width=0.3pt,->,out=30,in=150] (L) to (R);
            \draw[line width=0.3pt,->,out=30,in=150,transform canvas={yshift=5pt}] (L) to node[above] {$\widetilde\phi_i^\dagger \equiv B_i$} (R);
            \draw[line width=0.3pt,<-,out=-30,in=-150] (L) to (R);
            \draw[line width=0.3pt,<-,out=-30,in=-150,transform canvas={yshift=-5pt}] (L) to node[below] {$\phi_i^\dagger \equiv A_i$} (R);
	   \end{tikzpicture}
    \end{center}
    \caption{Description of the $U(N)_k\times U(N)_{-k}$ as a 3d $\mathcal N=2$ quiver gauge theory. The $(\phi_i,(\lambda_i)_\alpha)$ (and their `$tilde$' counterparts) denote an $\mathcal N=2$ chiral multiplet. There are two of each, which explains the double arrows. We have also included the standard notation for ABJM bi-fundamentals $A_i$ and $B_i$, which we will use throughout the paper and that are  defined in equation \eqref{eq:BPS_words}.}
    \label{fig:quiver_abjm}
\end{figure}
In the 3d $\mathcal N=2$ formalism, the ABJM theory can be described as a 3d $\mathcal N=2$ Chern-Simons matter theory with gauge group $U(N)\times U(N)$ and Chern-Simons levels $k$ and $-k$. This is conveniently described as a 3d $\mathcal N=2$ quiver gauge theory, as in Figure \ref{fig:quiver_abjm}, with gauge group $U(N)\times U(N)$.

The matter content consists of two bifundamental chiral multiplets containing four scalars:  $\phi_1,\phi_2,\widetilde\phi_1,\widetilde\phi_2$, and four fermions: $(\lambda_1)_\alpha,(\lambda_2)_\alpha,(\widetilde\lambda_1)_\alpha,(\widetilde\lambda_2)_\alpha$. Finally, there is the following superpotential term:
\begin{equation}
    \mathcal W = \frac{4\pi}{k}{\rm Tr}\left(\phi_1\widetilde\phi_1\phi_2\widetilde\phi_2-\phi_1\widetilde\phi_2\phi_2\widetilde\phi_1\right) ~.
\end{equation}
There are also two Chern-Simons gauge fields, and we review the explicit supersymmetry transformations in Appendix \ref{app:ABJM}.

Note that this description of the theory has a manifest $SU(2)\times SU(2)$ symmetry that rotates the two chiral multiplets. In particular $(\phi_1,\phi_2)$ are  $(\mbf{N},\overline{\mbf{N}})$ bifundamentals under the gauge group and transform as $(\mbf{2},\mbf{1})$ under this $SU(2)\times SU(2)$ symmetry. Whereas, $(\widetilde\phi_1,\widetilde\phi_2)$ are  $(\overline{\mbf{N}},\mbf{N})$ anti-bifundamentals that transform as $(\mbf{1},\mbf{2})$; the same is true for the corresponding fermions. It is perhaps interesting to note that these are arranged into the fundamental representations of $SU(4)_R$ as follows:  $(\phi_1,\phi_2,\widetilde\phi_1^\dagger,\widetilde\phi_2^\dagger)$ for the bosons and $(-(\lambda_2)_\alpha^\dagger,(\lambda_1)_\alpha^\dagger,-(\widetilde\lambda_2)_\alpha,(\widetilde\lambda_1)_\alpha)$ for the fermions. This $SU(4)_R$ can then be decomposed into the $U(1)_R$ symmetry that is manifest in the $\mathcal N=2$ Lagrangian and the `flavour' commutant $SU(2)\times SU(2)$. There is an additional $U(1)_t$ topological symmetry realised as a diagonal combination of the two $U(1)$ topological symmetries associated to the two gauge groups.\footnote{ The current for the $U(1)_T$ topological symmetry takes the form $*(F_1+F_2)$. Because of the Chern-Simons term, the equations of motion of the  gauge fields set to zero one combination of the topological symmetries and relate $k\,  U(1)_t$  to  the {\it baryonic symmetry} of the quiver. } This does not act on the fundamental scalars $\phi_i,\widetilde\phi_j$, but instead on monopole operators that we describe next.

\subsection{Monopole operators}
\label{subsec:monopole_review}

The non-perturbative physics of these three-dimensional theories is governed by monopole operators, which are local disorder operators creating quantised magnetic flux for the gauge fields. A 't Hooft monopole operator is defined by a Dirac monopole singularity for a $G$ gauge field at an insertion point
\begin{equation}
    A_{\pm} = \frac{\mbf{m}}{2}(\pm1-\cos \theta){\rm d}\phi ~,
\end{equation}
where $(\theta,\phi)$ are $S^2$ coordinates around the insertion, $\mbf m\in {\rm Lie}(G)$ is the magnetic charge, and $A_\pm$ are the corresponding connections on the northern/southern patch of the $S^2$. Single-valuedness of the transition function on the overall of the two patches imposes a (generalised) Dirac quantisation condition
\begin{equation}
    \exp\left(2\pi i \,\mbf m\right) = \mathbb 1_G ~,
\end{equation}
which implies the magnetic fluxes belong to the weight lattice of the GNO/Langlands dual group of $G$.

In the ABJM setting, this corresponds to a pair of magnetic fluxes  under the two $U(N)$ gauge groups, each specified by a set of $N$ integers,
\begin{equation} \mbf m=(m_1,m_2,\ldots , m_N)\, \qquad  \widetilde{\mbf m} =(\widetilde{m}_1, \widetilde{m}_2,\ldots , \widetilde{m}_N) \, .\end{equation}
The fluxes are only defined up to action of the Weyl group of $U(N)\times U(N)$, which acts by independently permuting the entries of $\mbf m$ and $\widetilde{\mbf m}$. 
Due to the Chern-Simons couplings, these so-called {\it bare} monopole operators, which we denote with  $T^{(\mbf m,\widetilde{\mbf m})}$, acquire electric charges under the two gauge groups. The monopole transforms in the $U(N)\times U(N)$ representation of highest weight $(-k\,\mbf m,k\,\widetilde{\mbf m})$, using notations where, if we order the integers as $m_1\ge m_2\ge \ldots m_N$ using the Weyl group action, $(k m_1, \ldots k m_N)$ is the $U(N)$ representation with Young tableau with rows of length $k m_i$ \cite{Klebanov:2008vq}. In particular the $U(1)\times U(1)$ charge of the monopole is $(- k \sum_i m_i, k \sum_j\widetilde m_j)$. A gauge invariant, {\it dressed} monopole operator is thus obtained by dressing the bare monopole operator with appropriate combinations of the bifundamental matter fields with compensating gauge charges.  

In ABJM, we have a further condition that $\sum_i m_i=\sum_j\widetilde m_j$, which ensures the presence of appropriate dressing fields in the residual theory to form gauge invariant monopole operators. This is a consequence of the diagonal gauge group $U(1)$ that does not act on the matter fields. 

The charge of the monopole operators under gauge and flavor symmetries is corrected by quantum effect \cite{Borokhov:2002cg,Benini:2011cma}. In particular, the $R$-charge (in $\mathcal{N}=2$ notation) of the bare monopole operator is given by:
\begin{equation}\label{Rchargemonopole}
    R(\mbf m,\widetilde{\mbf m}) = \sum_{i,j=1}^N \vert m_i-\widetilde m_j \vert - \sum_{1\leq i<j\leq N} \vert m_i-m_j \vert - \sum_{1\leq i<j\leq N} \vert \widetilde m_i-\widetilde m_j \vert ~,
\end{equation}
for magnetic fluxes $(\mbf m,\widetilde{\mbf m})$ for the $U(N)\times U(N)$ gauge group. Notice that $R(\mbf m,\widetilde{\mbf m}) =0$ if $\widetilde{\mbf m}=\mbf m$. Thus the simplest non-trivial dressed monopole operator appear roughly at R-charge $\mathcal O(k/2)$ corresponding to the $\widetilde{\mbf m}=\mbf m= (\pm1,0,\dots,0)$ sector dressed by the appropriate $k$ bifundamental scalars $\phi_i$ or $\widetilde\phi_j$. In particular, monopoles have large dimension and disappear from the spectrum in the limit $k\gg 1$.

\subsection{BPS cohomology}

We now describe the $1/12$th BPS sector of local operators in the $U(N)_k\times U(N)_{-k}$ ABJM theory. We choose the supercharges
\begin{equation}
    \mathcal Q=\left(Q_{34}\right)_- \;,\quad \mathcal S=\mathcal Q^\dagger=\left(S^{34}\right)^- ~. 
\end{equation}
Local operators invariant under the action of these supercharges satisfy 
\begin{equation}\label{eq:BPS_cond}
    \delta = \{\mathcal Q,\mathcal S\} = \mathcal D-R_4^4-R_3^3+M_-^- = \mathcal D-q_3-J \, \stackrel{!}{=} \, 0 ~,
\end{equation}
where $J$ is the Cartan generator associated to the spacetime $SO(1,2)$ algebra, and $q_3$ is  the third Cartan generator of $SO(2)^3\in SO(6)_R$. The full subalgebra preserved by the action of these supercharges is $OSp(4\vert2) \subset OSp(6\vert4)$, which corresponds to the algebra of an $\mathcal N=4$ supersymmetric quantum mechanics. By definition, local operators that are BPS under the action of $\mathcal{Q}$  can be organised into multiplets of $OSp(4\vert2)$. The commutant subalgebra is generated by the following set of supercharges that commute with $\mathcal Q$:
\begin{equation}
    \left(Q_{13}\right)_+ \;,\;\; \left(Q_{14}\right)_+ \;,\;\; \left(Q_{23}\right)_+ \;,\;\; \left(Q_{24}\right)_+ ~.
\end{equation}

It is easy to verify that the basic fields, henceforth referred to as the BPS letters, that satisfy condition \eqref{eq:BPS_cond} in the classical theory ($k\rightarrow\infty$) are:\footnote{For later convenience we have defined the fermionic letter in a way that swaps the $(12)$ label on the $\lambda,\widetilde\lambda$ fermions.}
\begin{equation}\label{eq:BPS_words}
    \begin{split}
        A_1 \coloneq \phi_1^\dagger \,,\; A_2 \coloneq \phi_2^\dagger \,&,\; B_1 \coloneq \widetilde\phi_1^\dagger \,,\; B_2 \coloneq \widetilde\phi_2^\dagger \;, \\[0.5mm]
        \psiA_1 \coloneq -(\lambda_2)_+ \,,\; \psiA_2 \coloneq (\lambda_1)_+ \,&,\; \psiB_1 \coloneq -(\widetilde\lambda_2)_+ \,,\; \psiB_2 \coloneq (\widetilde\lambda_1)_+ \;, \\[1mm]
        D \coloneq& D_{++}
    \end{split}
\end{equation}
where the precise definition of the derivative letter is clarified in Appendix \ref{app:ABJM}. The fact that the derivative $D$ is a BPS letter implies that arbitrary derivatives of the other BPS fields are still BPS. Although a component of the Chern--Simons gauge field satisfies the BPS condition \eqref{eq:BPS_cond}, it does not appear as an independent letter because it is not gauge invariant, and the corresponding field strength is not BPS, see Table \ref{tab:ABJM_charges}.  Note that none of the bifundamental BPS letters are gauge invariant on their own; however, gauge invariant `BPS words' can be formed by considering products of bifundamental and anti-bifundamental BPS letters, which furnish representatives of the corresponding BPS cohomologies. For later convenience, we define the following shorthand for the bosonic and fermionic BPS letters:
\begin{equation}
    Y_I \coloneq \{A_1,A_2,B_1,B_2\} \;,\; \Psi^I \coloneq \big\{\psiA_1,\psiA_2,\psiB_1,\psiB_2\big\} \;, \text{ where } I=1,2,3,4.
\end{equation}

In the large $k$ limit---or, equivalently, in the trivial monopole sector at any $k$---the BPS cohomology representatives contain an equal number of bifundamental and anti-bifundamental BPS letters to ensure gauge invariance, together with an arbitrary number of derivatives $D$. Thus, the basic building blocks in this sector are the products of one bifundamental and one anti-bifundamental, for instance ${\rm Tr}(A_1\psiA_2)$. Note that based on our re-definition, ${\rm Tr}(A_1\psiA_1) = {\rm Tr}(\phi_1^\dagger(\lambda_2)_+)$, which is indeed the right gauge-invariant product. This is no longer true in non-trivial monopole sectors, as the gauge charge of the product of many bifundamental letters can be cancelled by the electric charge of the bare monopole. For instance, in the $\mbf m=\widetilde{\mbf m}= (1,0,\dots,0)$ monopole sector, a gauge invariant word can take the form ${\rm Tr}\left(T^{(1,0,\dots,0)} B_1^k\right)$.\footnote{Whenever $\mbf m=\widetilde{\mbf m}$ we will abbreviate $T^{(\mbf m,\widetilde{\mbf m})}\equiv T^{\mbf m}$}.

We now present the explicit one-loop $\mathcal Q$ action on the BPS letters. The scalar fields remain BPS even with this $\mathcal Q$ action, and transform as
\begin{equation}
    \mathcal Q\,A_i = 0 \;,\quad \mathcal Q\,B_i = 0 ~.
\end{equation}
The $\mathcal Q$ action on the fermions picks up a $\mathcal O(1/k)$ term of the form
\begin{equation}\label{eq:Qaction_fermions}
    \mathcal Q\,\psiA_i = \frac{4\pi i}{k} \epsilon_{rs}B_rA_iB_s \;,\quad \mathcal Q\,\psiB_i = -\frac{4\pi i}{k} \epsilon_{rs}A_rB_iA_s ~.
\end{equation}
Finally, the $\mathcal Q$ action on the BPS letters with a $D$ acting on them can be expressed as\footnote{Note that here we use the convention $2\,O_{[r}\widetilde O_{s]} = \epsilon_{rs}\epsilon^{ab}O_a\widetilde O_b$.}
\begin{equation} \label{Qtransf}
    \begin{split}
        \mathcal Q\left(DA_i\right) &= \frac{8\pi}{k} \left( A_{[1}\psiA_{2]}A_i - A_i\psiA_{[2}A_{1]} + \psiB_{[1}B_{2]}A_i - A_iB_{[2}\psiB_{1]} \right) ~, \\
        \mathcal Q\left(DB_i\right) &= \frac{8\pi}{k} \left( \psiA_{[2}A_{1]}B_i - B_iA_{[1}\psiA_{2]} + B_{[2}\psiB_{1]}B_i - B_i\psiB_{[1}B_{2]} \right) ~, \\
        \mathcal Q\left(D\,\psiA_i\right) &= D\left(\mathcal Q\psiA_i\right) + \frac{8\pi}{k} \left( \psiA_{[2}A_{1]}\psiA_i + \psiA_iA_{[1}\psiA_{2]} + B_{[2}\psiB_{1]}\psiA_i + \psiA_i\psiB_{[1}B_{2]} \right) ~, \\
        \mathcal Q\left(D\,\psiB_i\right) &= D\left(\mathcal Q\psiB_i\right) + \frac{8\pi}{k} \left( A_{[1}\psiA_{2]}\psiB_i + \psiB_i\psiA_{[2}A_{1]} + \psiB_{[1}B_{2]}\psiB_i + \psiB_iB_{[2}\psiB_{1]} \right) ~.
    \end{split}
\end{equation}
\normalsize

We note here that BPS cohomologies are spanned by local operators, $\mathcal O$, that are $\mathcal Q$ closed, but not $\mathcal Q$ exact, {\it i.e.},
\begin{equation}
    \mathcal Q\,\mathcal O = 0 \quad\text{($\mathcal Q$ closed)} \;,\quad \mathcal O \neq \mathcal Q\,\mathcal \Lambda \quad\text{(not $\mathcal Q$ exact)} ~.
\end{equation}
As a result, two operators $\mathcal O$ and $\mathcal O'$ live in the same $\mathcal Q$ cohomology class if they are both $\mathcal Q$ closed and differ by a $\mathcal Q$ exact term. Since we are primarily interested in BPS cohomologies, it is important to remember that the local operators that we write down are only representatives of the BPS cohomology classes and not necesarily the exact quantum BPS operators. While BPS cohomologies have nice properties as couplings are varied, the actual BPS operators will be very coupling dependent.

\subsection{Superconformal Index}
\label{subsec:superconf_index}

The superconformal index of the ABJM theory quantised on $S^2\times\mathbb R$ is the Witten index
\begin{equation}
    \mathcal I(x;y_1,y_2) = {\rm Tr} \,(-1)^F\,x^{\mathcal D+J}\,y_1^{q_1}y_2^{q_2} ~,
\end{equation}
where as usual $\mathcal D, J$ are the dilation and $SO(1,2)$ Cartan generator operators that furnish the scaling dimension and Lorentz spin, respectively. On the other hand, $q_1$ and $q_2$ are two of the Cartan generators of $SO(6)_R$, distinct from the one that is used to define the BPS cohomology in \eqref{eq:BPS_cond}. Note that  the index is defined with respect to the supercharge $\mathcal Q$, and hence it only receives contributions from states annihilated by $\mathcal Q$ and its Hermitian conjugate $\mathcal S$; it thus exactly counts the BPS cohomology classes of interest.

The Cartan of the $SU(2)\times SU(2)$ flavour symmetry rotating the $\mathcal N=2$ chiral multiplets is realised as the diagonal and anti-diagonal combinations of $q_1$ and $q_2$. Therefore, redefining the fugacities as
\begin{equation}
    u = \sqrt{\frac{y_1}{y_2}} \;,\quad v = \sqrt{y_1y_2} ~,
\end{equation}
we can rewrite the superconformal index as
\begin{equation}\label{eq:index}
    \mathcal I(x;u,v,t_1,t_2) = {\rm Tr} \,(-1)^F\,x^{\mathcal D+J_3}\,u^{q_1-q_2}\,v^{q_1+q_2}\,t_1^{\mathcal T_1}\,t_2^{\mathcal T_2} ~,
\end{equation}
with $u$ and $v$ now allowing for a refined counting of operators that fit into the $SU(2)_u\times SU(2)_v$ flavour symmetry that rotates the chiral multiplets. We have also added additional fugacities $t_1,t_2$ which refine the index by the $U(1)$ topological charges associated to the two $U(N)$ gauge groups. As we have already discussed, the monopole fluxes in ABJM satisfy $\sum_i m=\sum_i\widetilde m$ and thus only one combination of these $U(1)$ symmetries will form the actual topological symmetry, whose fugacity is given by $t=t_1t_2$. We can directly see this from the index as we will explain shortly.

The superconformal index can be expressed as an integral over the two $U(N)$ gauge groups
\begin{equation}\label{eq:general_superconf_index}
    \begin{split}
        \mathcal I(x;u,v,t_1,t_2) &= \sum_{\mbf m,\widetilde{\mbf m}\,\in\,\mathbb Z^N} \frac{1}{(N!~)^2} \oint_{\mathbb T^N} \prod_{i=1}^N\frac{{\rm d}z_i}{2\pi iz_i} \oint_{\mathbb T^N} \prod_{i=j}^N\frac{{\rm d}w_j}{2\pi iw_j} \\
        &\qquad \times \prod_{i=1}^N z_i^{km_i}t_1^{m_i}\prod_{j=1}^N w_i^{-k\widetilde m_j}t_2^{\widetilde m_j} \; \mathcal I_{\rm vec}\left(x;\{z_i\},\mbf m\right)\,\mathcal I_{\rm vec}\left(x;\{w_j\},\widetilde{\mbf m}\right) \\
        &\qquad\quad \times \prod_{i,j=1}^N \mathcal I_{\rm chir}\left(x;u^{-1}z_i/w_j,m_i-\widetilde m_j\right) \, \mathcal I_{\rm chir}\left(x;v^{-1}w_i/z_j,\widetilde m_i-m_j\right) \\
        &\qquad\qquad\quad \times\; \mathcal I_{\rm chir}\left(x;u\,z_i/w_j,m_i-\widetilde m_j\right) \, \mathcal I_{\rm chir}\left(x;v\,w_i/z_j,\widetilde m_i-m_j\right) ~,
    \end{split}
\end{equation}
where $\mathcal I_{\rm vec}$ is the contribution of the $U(N)$ $\mathcal N=2$ vector multiplet,
\begin{equation}
    \begin{split}
        \mathcal I_{\rm vec}\left(x;\{z_i\},\mbf m\right) = \prod_{1\leq i<j\leq N} x^{-|m_i-m_j|}\, &(1-(-1)^{m_i-m_j}x^{|m_i-m_j|}z_i/z_j) \\
        &\times (1-(-1)^{m_i-m_j}x^{|m_i-m_j|}z_j/z_i) ~,
    \end{split}
\end{equation}
while $\mathcal I_{\rm chir}$ is the contribution of a single 3d $\mathcal N=2$ chiral multiplet
\begin{equation}
    \mathcal I_{\rm chir}(x;z,m) = (x^{\frac12}z^{-1})^{\frac{|m|}{2}}\prod_{i=0}^\infty  \frac{1-(-1)^mx^{\frac32+2i+|m|}z^{-1}}{1-(-1)^mx^{\frac12+2i+|m|}z} ~.
\end{equation}

\begin{table}[!h]
    \begin{center}\label{table_charges}
        \begin{tabular}{|c|c|c|c|c|c|} 
            \hline
            $\mathcal D+J_3$ & Field & Gauge $U(1)\times U(1)$ & $SU(2)_u$ & $SU(2)_v$ & $U(1)_t$ \\
            \hline\hline
            0 & $T^{(\mbf m,\mbf m)}$ & $(-k\sum_im_i,k\sum_im_i)$ & $0$ & $0$ & $\sum_im_i$ \\
            \hline
            \multirow{4}{*}{$\frac12$} & $A_1$ & $(-1,1)$ & $-1$ & $0$ & $0$ \\
            & $A_2$ & $(-1,1)$ & $1$ & $0$ & $0$ \\
            & $B_1$ & $(1,-1)$ & $0$ & $-1$ & $0$ \\
            & $B_2$ & $(1,-1)$ & $0$ & $1$ & $0$ \\
            \hline
            2 & $D$ & $(0,0)$ & $0$ & $0$ & $0$ \\
            \hline
            \multirow{4}{*}{$\frac32$} & $\psiA_1$ & $(1,-1)$ & $-1$ & $0$ & $0$ \\
            & $\psiA_2$ & $(1,-1)$ & $1$ & $0$ & $0$ \\
            & $\psiB_1$ & $(-1,1)$ & $0$ & $-1$ & $0$ \\
            & $\psiB_2$ & $(-1,1)$ & $0$ & $1$ & $0$ \\
            \hline
        \end{tabular}
    \end{center}
    \caption{Various fields that contribute to the index and their charges under various symmetries. Note that we only present the gauge charges under the diagonal $U(1)\times U(1)\subset U(N)\times U(N)$. The precise statement is that $B,\psiA$ are $(\mbf N,\overline{\mbf N})$ bifundamentals fields, while $A,\psiB$ are $(\overline{\mbf N},\mbf N)$ anti-bifundamental fields. Furthermore, the generic bare monopoles do not have vanishing $\mathcal D+J_3$ quantum numbers, however we will only encounter the $\mbf m=\widetilde{\mbf m}$ monopoles in this work, which do indeed have vanishing $\mathcal D+J_3$ quantum numbers.} \label{tablecharge_sec2}
\end{table}

Some comments are in order after the full gory expression for the superconformal index. First, notice that there is an overall sum over all flux sectors, and in each flux sector, the integral projects onto gauge invariants. Second, the index has been written assuming generic flux sectors through the two gauge groups. However, there is a decoupled $U(1)$ fugacity, let us call it $z_{\rm dec}$, and the corresponding integral simply reduces to
\begin{equation}
    \oint_{|z_{\rm dec}|=1} \frac{{\rm d}z_{\rm dec}}{2\pi i z_{\rm dec}} z_{\rm dec}^{\sum_i m_i-\sum_j \widetilde m_j} = \delta_{\sum_i m_i-\sum_j \widetilde m_j} ~,
\end{equation}
the same condition for magnetic fluxes that we noted in Subsection \ref{subsec:monopole_review}. Finally, this condition implies that the dependence on $t_1$ and $t_2$ in \eqref{eq:general_superconf_index} simplifies to
\begin{equation}
    \prod_{i=1}^N t_1^{m_i} \prod_{j=1}^N t_2^{\widetilde m_j} \delta_{\sum_i m_i-\sum_j \widetilde m_j} = (t_1 t_2)^{\sum_i m_i} \eqcolon t^{\sum_i m_i} ~,
\end{equation}
where the new fugacity $t=t_1t_2$ corresponds the $U(1)_t$ topological symmetry. As is evident, the topological charge in a given monopole sector is simply given by the sum of the magnetic flux components.

Thus, the superconformal index that we consider in this paper takes the form
\begin{equation}
    \mathcal I(x;u,v,t) = {\rm Tr} \,(-1)^F\,x^{\mathcal D+J_3}\,u^{q_1-q_2}v^{q_1+q_2}\,t^{\mathcal T} ~,
\end{equation}
with the integral \eqref{eq:general_superconf_index} and the simplification described above. This can be expanded as a power series in the variable $x$, with each coefficient being a linear combination of characters of the $SU(2)_u\times SU(2)_v\times U(1)_t$ global symmetry. For convenience, we present the various charges of all the BPS letters in Table \ref{tablecharge_sec2}. In the case of $k=1,2$, there is an enhancement of the $R$-symmetry to $SO(8)_R$. Thus, the $U(1)$ $R$-symmetry subalgebra used to define the index now has an $SU(4)\cong SO(6)$ commutant inside $SO(8)$ that corresponds to a flavour symmetry from the point of view of the index. This $SU(4)$ symmetry mixes operators across different monopole sectors at a fixed order of $x$. The $SU(4)$ symmetry breaks down to  $SU(2)_u\times SU(2)_v\times U(1)_t$ at generic $k$ and to $SU(2)_u\times SU(2)_v$ in the zero monopole sector, corresponding to the large $k$ limit.

The superconformal index of ABJM and related theories has been expanded at low order and organized into representations of the symmetry algebra in many different context, see for example \cite{Beratto:2021xmn,Hayashi:2022ldo}.  In the following we will expand it for the case of interest for our search of fortuity  and we apologize in advance if some expression has already appeared in the literature and we give no proper reference.

\subsection{A non-renormalization conjecture}

One of the main differences between ABJM and $\mathcal{N}=4$ SYM is the fact that ABJM does not have a continuous coupling: the levels of the Chern-Simons terms are quantized to be integers. This means that even in the type IIA limit, the string length in AdS units is not a continuous parameter of the theory, but is actually quantized. This implies additional subtleties for the classification of cohomologies, since one would like to make statements about black holes, which are relevant only when the theory is strongly coupled.

In the case of $\mathcal{N}=4$ SYM, this is addressed in terms of a non-renormalization conjecture:\footnote{There is tension with this conjecture as has been observed for gauge groups other than $SU(N)$ in \cite{Chang:2025mqp,Choi:2025bhi}.} it is conjectured in \cite{Grant:2008sk} that the operators that are BPS at one-loop, are actually BPS at finite coupling. The conjecture implies that there is a huge lifting of operators between the free theory and the strongly coupled theory, but that the entirety of this lifting is realized already at one-loop.

Here, we are in a different situation since there is no continuous coupling. As we have just seen in the previous section, the superconformal index explicitly depends on $k$, so there is clearly no sense in which we can have a non-renormalization conjecture as powerful as that of $\mathcal{N}=4$ SYM. Nevertheless, we have seen that the superconformal index depends on $k$ only in the sense that the effect of monopole sectors changes as $k$ changes. In particular, the lowest dimension state of any non-trivial monopole sector arises at order $x^{k/2}$. The index is therefore insensitive to $k$ for all operators contributing to the term $x^n$ with $n<k/2$. This enables us to formulate a one-loop exactness conjecture for ABJM.

\begin{theorem}
We conjecture that the BPS cohomologies of ABJM for operators with $\Delta<k/2$, at fixed $\Delta$ but for varying $k$ are given exactly by the cohomologies that are closed under the one-loop supercharge, as long as $k/2>\Delta$.
\end{theorem}

It worth noting that at $k=\infty$, there are extra BPS cohomologies, which is manifest from the form of the supercharge transformation equations \eqref{Qtransf}. So just like in $\mathcal{N}=4$ SYM, there is a big change between the free and one-loop cohomologies, and we conjecture that the one-loop exactness holds in ABJM as well, with the extra caveat that the quantum numbers of the state lower-bound the value of $k$ up to which this conjecture is meant to hold. Non-perturbatively in $k$, we know monopole operators have an important effect, and as we will see, can both add or remove cohomologies.

It is also important to note that even if true, this conjecture does not directly impact the spectrum of black holes once we take a holographic limit. Black holes appear at $\Delta\sim N^{3/2}$ or $N^2$ in the M-theory or type IIA limit, which is always in a regime where $\Delta$ fails to satisfy $\Delta<k/2$. Therefore, if true, the non-renormalization conjecture is very powerful but not as powerful as it is in $\mathcal{N}=4$ SYM, where it also impacts the black hole spectrum. 

\subsection{Monotones vs Fortuitous cohomologies}

Before we dive into the details of the BPS cohomologies in ABJM, let us review the definition of monotone and fortuitous states \cite{Chang:2024zqi}. At a given $N$, monotone cohomologies are those that are BPS and continue to be BPS as $N$ is increased. Therefore, they do not rely on trace relations to be BPS. A monotone operator $O_M$ thus satisfies 
\be
\mathcal{Q} O_M= 0 \,, \quad \forall N \,.
\ee
The conjecture of \cite{Chang:2024zqi} is that such operators fully span the set of all graviton and multi-graviton states. At infinite $N$, these converge to the spectrum of free KK fields and to the associated Fock space. Note that the spectrum of monotone operators is \textit{not} $N$-independent. At finite $N$, the spectrum is different from the infinite $N$ theory. This is because certain single-trace operators can be reduced to multi-trace operators due to trace relations and are thus excluded from the spectrum. Moreover, certain multi-trace operators become $\mathcal{Q}$-exact at finite $N$ due to trace relations, and must therefore be excluded from the cohomologies. The finite $N$ monotone cohomologies are thus a strict subset of cohomologies at higher $N$, since the trace relations are strictly smaller as $N$ is increased.

In the spirit of \cite{Chang:2024zqi} we thus define the graviton finite $N$ spectrum as the set of all monotone operators at given value of $N$.
One may be tempted to define the graviton finite $N$ spectrum as the set of all cohomologies built from products of BPS {\it single-trace operators}, up to taking into account trace relations. In all example studied so far, this is equivalent to the set of operators that are BPS at all values of $N$. In this paper, we will see that the two definitions are actually not equivalent, as we find concrete examples of single-trace non-graviton BPS cohomologies, that are BPS at very low values of $N$ (at $N=1$). So for the rest of the paper, we will take the monotone definition to be the BPSness of operators as $N$ is varied, and not what is generated by the single-trace BPS operators.

On the other hand, fortuitous operators are all the cohomologies that are not of the graviton type. They should therefore correspond to black hole operators. The conjecture of \cite{Chang:2024zqi} is that they are all of the following form: the action of the supercharge on such a fortuitous operator $O_F$ reads
\be
\mathcal{Q}\, O_F = \text{trace relation at } N=N^* \,.
\ee
Therefore, the operator is fortuitous at $N=N^*$, and eventually, at some $N^0>N^*$, the operator fails to be BPS as the trace relations have weakened. Our goal will be to find operators of this type in ABJM theory.

\section{Fortuity in $U(1)_k \times U(1)_{-k}$ ABJM \label{sec:3}}

In this section, we address the problem of fortuity in the $U(1)_k\times U(1)_{-k}$ theory. We begin by noting that in the $U(1)_k\times U(1)_{-k}$ ABJM theory, all the BPS letters are simply $1\times 1$ matrices. This implies that the $\mathcal Q$ action in \eqref{eq:Qaction_fermions} vanishes yielding
\begin{equation}\label{eq:Qaction_U1}
    \mathcal Q\,A_i = 0 \;,\quad \mathcal Q\,B_i = 0 \;,\quad \mathcal Q\,\psiA_i = 0 \;,\quad \mathcal Q\,\psiB = 0 ~.
\end{equation}
The property that all the BPS letters are simply $1\times 1$ matrices also implies that there are no non-trivial trace relations in this theory.\footnote{But we will see that even the trivial trace relations will have drastic consequences in terms of fortuity.} Using this remarkable simplification in the $\mathcal Q$ action, we observe that there are plenty of states that are BPS 
only at $N=1$, which, according to the definition, will be called fortuitous. This give a large  number of fortuitous BPS cohomologies in the $N=1$ ABJM theory for $k>1$. In the special case $N=1$ and $k=1$ the index is freely generated and there is no fortuity. We also perform a direct calculation of the index for general $k$ which allows us to enumerate all cohomologies  in the $N=1$ theory.

\subsection{Large $k$}

 We first review the superconformal index for the $U(1)_k\times U(1)_{-k}$ ABJM index at large $k\gg1$. As explained in Section \ref{subsec:monopole_review}, the first non-trivial monopole background BPS cohomology appears at order $k/2$. Therefore, the index at large $k$ is equivalent to the trivial $m=\widetilde m=0$ magnetic flux sector.\footnote{By a slight abuse of terminology, we will call the zero monopole sector the large $k$ index, but this is only strictly speaking true for the expansion of the index up to $x^{k/2}$. Since we are mostly concerned with the first few orders of the expansion, this will be sufficient.} Using \eqref{eq:general_superconf_index}, this index can thus be expressed as
\begin{equation}\label{eq:U1_bareindex}
    \begin{split}
        \mathcal I_{N=1,k\gg1}(x;u,v) = \oint \frac{{\rm d}z}{2\pi iz} \oint \frac{{\rm d}w}{2\pi iw} &\prod_{n=0}^\infty \frac{\big(1-u^{-1}\,x^{\frac32+2n}\frac{z}{w}\big)\big(1-ux^{\frac32+2n}\frac{z}{w}\big)}{\big(1-u^{-1}\,x^{\frac12+2n}\frac{z}{w}\big)\big(1-ux^{\frac12+2n}\frac{z}{w}\big)} \\
        &\quad \times \frac{\big(1-v^{-1}\,x^{\frac32+2n}\frac{w}{z}\big)\big(1-vx^{\frac32+2n}\frac{w}{z}\big)}{\big(1-v^{-1}\,x^{\frac12+2n}\frac{w}{z}\big)\big(1-vx^{\frac12+2n}\frac{w}{z}\big)} ~.
    \end{split}
\end{equation}
Note that there is no fugacity for the $U(1)$ topological symmetry as we are in the zero monopole sector. As evident from the expression, the integrand only depends on the ratio $z/w$, and is independent of $zw$, the latter being the fugacity for the decoupled $U(1)_{\rm dec}$.

 In order to understand the structure of the co-homology and the presence or absence of fortuity, we start by listing operators that contribute to the index at low orders. It is also useful to organize the index as an expansion in characters of the symmetry $SU(2)_u\times SU(2)_v$. Then the index expanded to low orders reads:
\begin{equation} \label{neq1largekindex}
    \begin{split}
        \mathcal I_{N=1,k\gg1}(x;u,v) = 1 &+ \big(\chiT{2}(u)\chiT{2}(v)\big)\,x + \big(\chiT{3}(u)\chiT{3}(v) - \chiT{3}(u) - \chiT{3}(v) -2 \big) \, x^2 \\
        &+ \big(\chiT{4}(u)\chiT{4}(v) + \chiT{2}(u)\chiT{2}(v) + \underline{2\chiT{2}(u)\chiT{2}(v)} \\
        & - \chiT{2}(u)\chiT{4}(v) - \chiT{4}(u)\chiT{2}(v) - \underline{2\chiT{2}(u)\chiT{2}(v)}\big)\,x^3 + \mathcal O(x^4) ~,
    \end{split}
\end{equation}
where $\chi_{\mbf\lambda}$ is character of the $SU(2)$ representation of dimension $\lambda$
\begin{equation}
    \chi_\lambda(w) = \frac{w^\lambda-w^{-\lambda}}{w -w^{-1}} =\sum_{k=-(\lambda-1)}^{\lambda-1} w^k 
\end{equation}
The underlined terms cancel each other and they have been added by hand to match the set of operators that we count at this level. This is a general feature of the index, since non-trivial boson/fermion cancellations can occur due to the $(-1)^F$ term in the index. We now list the cohomological representative for each of these terms. We use a notation with parenthesis that represent ``traces". Of course, these are not real traces as the $N=1$ theory has no matrices. But the parenthesis label gauge-invariant sub-blocks, and have natural embeddings into the higher $N$ theories. We will see that there are subtleties with this prescription, related to the nature of the trace relations at $N=1$.

\vspace{1.5em}

\noindent{\bf Order $x$} : These are the four basic gauge-invariant mesons constructed out of the bosonic BPS letters. Schematically, we represent them as
\begin{equation}\label{eq:U1_AB}
    \SUSU{2}{2} :\, (AB) ~.
\end{equation}
These basic gauge-invariants are {\it single-trace} operators and survive at large $N$. They are just single graviton states. 

\vspace{1.5em}

\noindent{\bf Order $x^2$} : These are a total of 9 bosons and 8 fermions. The bosons are of the form 
\begin{equation}\label{eq:U1_ABAB}
    \SUSU{3}{3} :\, (AB)(AB) ~.
\end{equation}
and are multi-gravitons corresponding  to products of the mesons $(AB)$.  They transform in the $(\mbf 3,\mbf 3)$ since  the BPS letters all commute with each other. We could have expected $10$ states of the form $(AB)(AB)$.\footnote{The $6$ antisymmetric combinations clearly vanish.} However the four mesons $M_{ij}=(A_iB_j)$ are not independent:  they satisfy the relation $M_{11}M_{22}=M_{12}M_{21}$ since  the fields commute. This is the general structure of the trace relations at $N=1$. The fermionic contribution at this order contains $(A\psiA)$ and $(B\psiB)$, which are both in $\mbf 3+\mbf1$ of the respective $SU(2)$ flavour symmetries,
\begin{equation}
    \SUSU{3}{1} + \SUSU{1}{1} :\, (A\psiA) \;, \qquad\; \SUSU{1}{3} + \SUSU{1}{1} :\, (B\psiB) ~.
\end{equation}
The split is simply due to the symmetrisation and anti-symmetrisation of the indices. On general grounds \cite{Cordova:2016emh}, at order $x^2$ we expect to find fermionic contributions from multiplets associated with the global symmetries of the theory that commute with $\mathcal Q$. We indeed find states in the adjoint of $SU(2)_u\times SU(2)_v\times U(1)_t \times U(1)_{o}$ where $U(1)_o$ is the {\it flavour} symmetry that rotates the four chiral multiplets with the same charge. This symmetry  exists only for $N=1$, since the superpotential is zero. The current corresponding to $U(1)_o$
\begin{equation}
    \psi_0 = \epsilon_{ij}\left ( A_i\psi^A_j +B_i\psi^B_j \right )
\end{equation}
thus represents the first example of a single-trace operator that is not a graviton. It is also our first example of fortuity. 
For general $N$, $\mathcal Q \psi_0=  \frac{8\pi i}{k} \epsilon_{ij}\epsilon_{rs} (A_iB_rA_jB_s)$  and the right-hand side is different from zero if $N>1$.

\vspace{1.5em}

\noindent{\bf Order $x^3$} : At this order, we have to add and subtract some representations in the index to find agreement with the explicit operators we can count, see equation \eqref{neq1largekindex}. First, we find bosonic multi-gravitons and descendents of single gravitons 
\begin{equation}
    \SUSU{4}{4} :\, (AB)(AB)(AB) \;,\qquad\; \SUSU{2}{2} :\, D(AB) \;,\qquad \; \SUSU{2}{2} :\, (ADB) ~,
\end{equation}
where the third type of derivative term $([DA]B)$ is absent since it is a linear combination of the $D(AB)$ and $(ADB)$. Next, the fermionic elements are made out of products of the mesons at order $x$ and the fermions at order $x^2$
\begin{equation}
    \SUSU{4}{2} + \SUSU{2}{2} :\, (\psiA A)(AB) \;,\qquad\; \SUSU{2}{4} + \SUSU{2}{2} :\, (\psiB B)(AB) ~.
\end{equation} 
Finally, the bosons $\psiA\psiB$ live in the $(\mbf2,\mbf2)$ representation and account for the remaining term in the index expansion
\begin{equation}
    \SUSU{2}{2} :\, (\psiA\psiB) ~.
\end{equation}
Notice that the $(\mbf2,\mbf2)$ representations appear both in the bosonic and fermionic sector and they cancel out in the index.
Notice also that the operators $(ADB)$ and $(\psiA\psiB)$ are BPS individually only at $N=1$. As we will see, starting with $N=2$, only one linear combination of them is annihilated by $\mathcal Q$. The other combination is another example of a single-trace fortuitous operator. On the other hand, all 
$24$ fermionic operators can be written as linear combinations of multi-gravitons.\footnote{It is easy to see that even  $\psi_o (AB)$ is equivalent to a linear combination of product of gravitons due to the extreme trace relations.}
Following this simple method, one can construct the cohomological representatives at any order in $x$ in terms of the basic BPS letters.

Fortunately, we do not have to do this because we can understand the spectrum of BPS operators directly from the superconformal index \eqref{eq:U1_bareindex}. The integral over the decoupled $U(1)_{\rm dec}$ can be easily performed in \eqref{eq:U1_bareindex}, leaving behind the following integral in terms of $\gamma\coloneq z/w$,
\begin{equation}\label{eq:U1Largek_index}
    \begin{split}
        \mathcal I_{N=1,k>>1}(x) = \oint_{|\gamma|=1} \frac{{\rm d}\gamma}{2\pi i\gamma} &\prod_{n=0}^\infty \frac{\big(1-u^{-1}x^{\frac32+2n}\gamma\big)\big(1-u\,x^{\frac32+2n}\gamma\big)}{\big(1-u^{-1}x^{\frac12+2n}\gamma\big)\big(1-u\,x^{\frac12+2n}\gamma\big)} \\
        &\quad \times \frac{\big(1-v^{-1}x^{\frac32+2n}\gamma^{-1}\big)\big(1-v\,x^{\frac32+2n}\gamma^{-1}\big)}{\big(1-v^{-1}x^{\frac12+2n}\gamma^{-1}\big)\big(1-v\,x^{\frac12+2n}\gamma^{-1}\big)} ~.
    \end{split}
\end{equation}
This is just counting all the words  made out of bifundamental/anti-bifundamental BPS letters and derivatives with the constraint that the words must be gauge invariant. Indeed, it is easy to recognise the integrand in \eqref{eq:U1Largek_index} as the contribution of the BPS letters listed in \eqref{eq:BPS_words} (bosons with charge $1/2$ and fermions with charge $3/2$) with the appropriate $SU(2)_u\times SU(2)_v$ charges as given in Table \ref{tablecharge_sec2}. The product over $n$ represents the action of the BPS derivative $D$, and finally the integral imposes gauge invariance. It would be an interesting exercise to extract the fortuitous partition function from this, by carefully removing all multi-gravitons. 

\subsection{Finite $k$}

The problem at $N=1$ and finite $k$ retains the simplicity due to \eqref{eq:Qaction_U1}, however, as opposed to the large $k$ case, we now have to consider non-trivial monopole sectors. One of the main features of non-trivial monopole sectors is that the bifundamental and anti-bifundamental BPS letters no longer need to be paired. This is required in the trivial monopole sector because of gauge invariance; however, the bare monopole carries a gauge charge, so this argument no longer holds. Although this adds a significant amount of richness to the theory, we nevertheless demonstrate that the structure of fortuity is very similar to the large $k$ case, except at $k=1$.

As before, we begin by stating the superconformal index of the $U(1)_k\times U(1)_{-k}$ ABJM theory at finite $k$ by specialising\eqref{eq:general_superconf_index}\footnote{Here we are using the notation that $f(x^{\pm1})=f(x)f(x^{-1})$.}
\begin{equation}\label{eq:U1k2_index}
    \begin{split}
        \mathcal I_{N=1,k}(x;u,v,t=t_1t_2) = &\sum_{(m,\widetilde m)\in\mathbb Z^2} \oint \frac{{\rm d}z}{2\pi iz} \oint \frac{{\rm d}w}{2\pi iw}\,z^{km}w^{-k\widetilde m} t_1^mt_2^{\widetilde m} \\
        &\qquad \times \; \mathcal I_{\rm chir}\left(x;u^{\pm1}z/w,m-\widetilde m\right) \\
        &\qquad\; \times \; \mathcal I_{\rm chir}\left(x;v^{\pm1}w/z,m-\widetilde m\right) ~.
    \end{split}
\end{equation}
As discussed before, integrating out the overall $U(1)$ from the superconformal index imposes the condition $m=\widetilde m$ with a remaining integral over the variable $\gamma=z/w$. Doing this explicitly, we obtain the result\footnote{One must always be careful when performing such a parameter redefinition in the index in order to obtain the right Jacobian factor. The transformation is on the torus $|z|=|w|=1$ and the correct Jacobian factor can be found by carefully analysing the transformation of the corresponding unit cell.}
\begin{equation}
    \begin{split}
        \mathcal I_{N=1,k}(x;u,v,t) = \sum_{m\in\mathbb Z} \oint \frac{{\rm d}\gamma}{2\pi i\gamma}\,\gamma^{km} &t^m \,\prod_{n=0}^\infty \frac{(1-u^{-1}x^{\frac32+2n}\gamma)(1-u\,x^{\frac32+2n}\gamma)}{(1-u^{-1}x^{\frac12+2n}\gamma)(1-u\,x^{\frac12+2n}\gamma)} \\
        &\quad\times \, \frac{(1-v^{-1}x^{\frac32+2n}\gamma^{-1})(1-v\,x^{\frac32+2n}\gamma^{-1})}{(1-v^{-1}x^{\frac12+2n}\gamma^{-1})(1-v\,x^{\frac12+2n}\gamma^{-1})} ~.
    \end{split}
\end{equation}
Note that the only dependence on the magnetic flux $m$ is via the monomial $\gamma^{km}t^m$. This lets us evaluate the following sum explicitly, where $\gamma=e^{2\pi i\zeta},t=e^{2\pi i\theta}$:
\begin{equation}
    \sum_{m\in\mathbb Z}\gamma^{km}t^m = \sum_{m\in\mathbb Z} e^{2\pi ikm(\theta+\zeta/k)} = \frac{1}{k}\,\sum_{l=0}^{k-1}\delta\left(\theta+\frac{\zeta-l}{k}\right) ~.
\end{equation}
Substituting this back into the index, we obtain
\begin{equation}
    \begin{split}
        \mathcal I_{N=1,k}(x;u,v) = \frac{1}{k}\,\sum_{l=0}^{k-1}&\prod_{n=0}^\infty \frac{(1-u^{-1}x^{\frac32+2n}t^{-1/k}e^{2\pi i\frac{l}{k}})(1-u\,x^{\frac32+2n}t^{-1/k} e^{2\pi i\frac{l}{k}})}{(1-u^{-1}x^{\frac12+2n}t^{-1/k}e^{2\pi i\frac{l}{k}})(1-u\,x^{\frac12+2n}t^{-1/k}e^{2\pi i\frac{l}{k}})} \\
        &\quad \times \, \frac{(1-v^{-1}x^{\frac32+2n}t^{1/k}e^{-2\pi i\frac{l}{k}})(1-v\,x^{\frac32+2n}t^{1/k}e^{-2\pi i\frac{l}{k}})}{(1-v^{-1}x^{\frac12+2n}t^{1/k}e^{-2\pi i\frac{l}{k}})(1-v\,x^{\frac12+2n}t^{1/k}e^{-2\pi i\frac{l}{k}})} ~,
    \end{split}\label{indexfinitek}
\end{equation}
where we have evaluated the integral using the delta function.

In  general, we expect to find all the fortuitous states we found for large $k$, since the zero monopole sector is a proper subset of the spectrum at finite $k$, plus new ones made with monopole operators. This will be true for generic $k$, with the exception of $k=1$ where want to argue that there are no-fortuitous states at all. Although it is difficult to compute the co-homology and the very notion of relations is not clear in the presence of monopole operators, we can observe that the index for $k=1$ seems to account for  multi-graviton BPS states only. For $k=1$ indeed we find the index of a free theory consisting of four chiral fields.  Each of the four bosonic BPS letters $Y_I=(A_1,A_2,B_1,B_2)$ and the four fermionic letters $\Psi^I =(\psi^A_1,\psi^A_2,\psi^B_1,\psi^B_2)$ can be made gauge invariant by combining them with monopoles $T^{\pm 1}$ which carry electric charge $\pm (-1,1)$
under the gauge group $U(1)\times U(1)$. These operators survive at $N=\infty$ and are single graviton states.
The index \eqref{indexfinitek} for $k=1$ just counts all possible products of these operators and the BPS derivative $D$, and thus can be seen as a multi-graviton index. The reader may wonder what happened to the fortuitous we found in the zero monopole sector.
The crucial point is that all these operators can be now written as products of single gravitons, since each letter now can be gauge invariant by itself.

Things are more complicated for $k>1$.
The BPS states  are now a  $\mathbb{Z}_k$ projection of the BPS states for $k=1$.
This is manifest from the index \eqref{indexfinitek} which is a $\mathbb{Z}_k$ average of the $k=1$ index. This is due to the fact that the monopoles $T^{\pm m}$ now carry electric charge $\pm (-k m,k m)$ and need to be dressed with products of BPS letters $Y_I$ and $\Psi_i $ and their derivatives with total opposite charge.
But a quick inspection of the charges of the fields reveals that these are precisely the products of letters invariant under the $\mathbb{Z}_k$ action
\begin{equation}\label{Zk}
A,\psiB \rightarrow e^{\frac{2\pi i}{k}} A, \psiB\, ,\qquad B,\psiA \rightarrow e^{-\frac{2\pi i}{k}} B, \psiA\, ,\end{equation}
combined with derivatives. For $k>1$, the current $J_o$ and all the other fortuitous we found at large $k$ are single-trace operators and they disappear from the spectrum at $N=\infty$. Other fortuitous operators come from sectors with non-vanishing monopole number. Notice that the BPS co-homology cannot be expressed as a collection of multi-gravitons states but can be obtained by taking arbitrary products of the {\it single-trace} generators 
\begin{equation}
    (AB)\, ,\quad (A\psi^A)\, ,\quad (B\psi^B)\, ,  \quad  (T^{-1} A^k) \, ,  \quad  (T^{1} B^k) \, ,  \quad  (T^{-1} \psi_B^k) \, ,  \quad  (T^{-1} \psi_A^k) \, , 
\end{equation}
and similar ones including derivatives.  The cohomology is no longer  freely generated. For example, the  bosonic generators
$M=(AB)$, $\tau=(T^{-1} A^k)$ and $\bar\tau=(T^{+1} B^k)$ are subject to constraints of the form $M M=M M$ and $M^k=\tau \bar\tau$, and similarly for fermions. 

For concreteness, we present the first few terms in the expansion of the index for $k=2$, as the exposition for $N=1$ will also aid the later discussion for the $N=2$ case. This is a special value of the Chern--Simons level, as there is a supersymmetry enhancement to $\mathcal N=8$, thus upgrading the flavour symmetry seen by the index from $SU(2)_u\times SU(2)_v\times U(1)_t$ to an $SU(4)_{a_i}$. Here, $a_i$ are the Cartan labels for the $SU(4)$ symmetry, and we work with the conventions where $a_1a_2a_3a_4=1$. Thus, the coefficients in the index can now be rearranged into representations of this $SU(4)$ flavour symmetry.

As before, we present the index explicitly at low orders, now written  in terms of the $SU(4)$ representations:
\begin{equation}
    \begin{split}
        \mathcal I_{N=1,k=2}(x;a_i) = 1 &+ \chiSU{10}(a_i)\,x + \Big( \chiSU{35}(a_i) - \chi^{SU(4)}_{\mbf{15}}(a_i) - 1 \Big)\,x^2 \\
        &+ \Big( \chiSU{84^{''}}(a_i) + 2\,\chi^{SU(5)}_{\mbf 6}(a_i) +\underline{\chiSU{10}(a_i)} \\[1mm]
        &\qquad\qquad\qquad\qquad - \, \chiSU{70}(a_i) - \underline{\chiSU{10}(a_i)} \Big)\,x^3 + \mathcal O(x^4) ~,
    \end{split}
\end{equation}
where the underlined terms have been added by hand to match the spectrum that we will enumerate next. Note that the index \eqref{eq:U1k2_index} generates a series in $x$ with coefficients that are Laurent polynomials in $u,v,t$. The map to the $SU(4)$ fugacities is given by
\begin{equation}\label{eq:SU4_map}
    a_1 = \frac{u}{t^{\frac12}} \;,\; a_2 = \frac{1}{u\,t^{\frac12}} \;,\; a_3 = v\,t^{\frac12} \;,\; a_4 = \frac{t^{\frac12}}{v} ~.
\end{equation}
It is clear that under this map, the fundamental of $SU(4)$ splits into $(\mbf 2,\mbf 1)_{-\frac12}+(\mbf 1,\mbf 2)_{\frac12}$ where we express the $SU(2)\times SU(2)$ representation in braces and the $U(1)$ topological charge in the subscript. The character can be expressed as
\begin{equation}
    \chi^{SU(4)}_{\mbf 4}
    (a_i) = a_1+a_2+a_3+a_4 = \left(u+\frac{1}{u}\right)\,t^{-\frac12} + \left(v+\frac{1}{v}\right)\,t^{\frac12} = \chiT{2}(u)\,t^{-\frac12}+\chiT{2}(v)\,t^{\frac12} ~.
\end{equation}

The letters $Y_I$ transform in the anti-fundamental representation of $SU(4)$ and the letters $\Psi^I$ in the fundamental. Although the letters are not gauge-invariant, we can form gauge invariants by using monopoles. We know from the index that the corresponding BPS operators must be invariant under the $\mathbb{Z}_2$ transformation \eqref{Zk}, which now simply changes simultaneously the sign of $Y_I$ and $\Psi^I$. The BPS operators thus contain an even number of letters $Y_I$ and $\Psi^I$.
 
\vspace{1.5em}

\noindent{\bf Order $x$} : The symmetric representation of $SU(4)$ decomposes as
\begin{equation}
    \overline{\mbf{10}} \longrightarrow \SUSU{3}{1}_{-1}+\SUSU{2}{2}_0+\SUSU{1}{3}_1 ~.
\end{equation}
The $(\mbf2,\mbf2)_0$ here corresponds to the zero flux sector, and hence we can match this result to the $(AB)$ states in \eqref{eq:U1_AB}. The $\pm1$ topological charges correspond to the $\mbf m=\widetilde{\mbf m}=\pm 1$ monopole sector, which has gauge charge $(\mp 2,\pm 2)$. This implies that the bare monopole is dressed by $B_1^2,B_1B_2,B_2^2$ or $A_1^2,A_1A_2,A_1^2$. Summarising, the operators that contribute at this order are
\begin{equation}\label{eq:U1_orderx}
    \SUSU{3}{1}_{-1} : \, (T^{-1}AA) \;,\; \SUSU{2}{2}_0 : \, (AB) \;,\; \SUSU{1}{3}_1 : \, (T^{1}BB) ~.
\end{equation}
We continue to write parenthesis here, to highlight gauge-invariant blocks, even if the notion of trace is a slight abuse. All of these operators can be compactly written as $(YY)$, which indeed transforms in the representation $\overline{\mbf{10}}$ because all letters commute. 

\vspace{1.5em}

\noindent{\bf Order $x^2$} : This order contains a single bosonic $\overline{\mbf{35}}$ dimensional $SU(4)$ representation and a total of 16 fermions split as $\mbf{15}+\mbf1$. The fermionic states account for the symmetry $SU(4)\times U(1)_o$ preserved by $\mathcal Q$,
where $U(1)_o$ is the accidental symmetry existing only for $N=1$. These representations decompose as
\begin{equation}
    \begin{split}
    \overline{\mbf{35}} \longrightarrow \SUSU{5}{1}_{-2} + \SUSU{4}{2}_{-1} + \SUSU{3}{3}_0 + \SUSU{2}{4}_1 + \SUSU{1}{5}_2 ~, \\
        \mbf{15} \longrightarrow \SUSU{2}{2}_{-1} + \SUSU{3}{1}_0 + \SUSU{1}{1}_0 + \SUSU{1}{3}_0 + \SUSU{2}{2}_1 ~.
    \end{split}
\end{equation}
The $(\mbf3,\mbf3)$ in the zero topological charge sector are the $9$ $(AB)(AB)$ presented in \eqref{eq:U1_ABAB}. The remaining bosons are dressed monopole operators, 
\begin{equation}
    \begin{split}
        \SUSU{5}{1}_{-2} &: \, (T^{-2}AA)(AA) \;,\; \SUSU{4}{2}_{-1} : \, (T^{-1}AA)(AB) \;,\; \SUSU{3}{3}_0 : \, (AB)(AB) ~, \\
        \SUSU{2}{4}_1 &: \, (T^{1}BB)(AB) \;,\; \SUSU{1}{5}_2 : \, (T^{2}BB)(BB) ~,
    \end{split}
\end{equation}
while the fermions are simply arranged as follows
\begin{equation}
    \begin{split}
        \SUSU{2}{2}_{-1} :\, (T^{-1}A\psiB) \;&,\; \SUSU{2}{2}_1 :\, (T^{1}B\psiA) ~, \\
        \SUSU{3}{1}_0+\SUSU{1}{1}_0 :\, (A\psiA) \;&,\; \SUSU{1}{3}_0+\SUSU{1}{1}_0 :\, (B\psiB)
    \end{split}
\end{equation}
Note that for the bosons, we picked a particular way to define gauge-invariant block, but this way is not unique, due to the trace relations at $N=1$. Above, we picked a particular representative of this. More compactly, we have $\overline{\mbf{35}}$ bosonic $(YY)(YY)$ and $\mbf{15}+\mbf 1$ fermionic $(Y\Psi)$, where the representation under $SU(4)$ follows from the fact that all letter commutes. As we discussed above, the fermionic singlet is a fortuitous state. 

\vspace{1.5em}

\noindent{\bf Order $x^3$} : At this order, there is once again a cancellation of full $SU(4)$ representations between bosons and fermions. Therefore, the exact operators can only be matched with the index counting after they are explicitly added. Instead of decomposing the $SU(4)$ representations and identifying the operators in the various $SU(2)\times SU(2)$ representations explicitly, we use the schematic notation where we count the representations using $X_I$ and $\Psi^I$. The bosonic operators are 
\begin{equation}
    \overline{\mbf{84}^{''}} :\, (YY)(YY)(YY) \;,\; \overline{\mbf{10}} :\,  D(YY) \;,\; \mbf 6 :\, (YDY) \;,\; \mbf 6 :\, (\Psi\Psi) ~,
\end{equation}
and the fermionic are
\begin{equation}
    \overline{\mbf{70}} + \overline{\mbf{10}} :\, (\Psi Y)(YY) ~.
\end{equation}
These can be decomposed in  $SU(2)\times SU(2)$ representations and monopole sectors as we did at orders $x$ and $x^2$.
One linear combination of the two $\overline{\mbf 6}$ is the $SU(4)$-completion of fortuitous states we found for large $k$  and we expect them to be fortuitous.

\section{Fortuity in $U(2)_k \times U(2)_{-k}$ at large $k$ \label{sec:4}}

In this section, we study fortuity in the weakly-coupled $U(2)_k\times U(2)_{-k}$ ABJM theory, {\it i.e.}, in the $k\gg1$ limit. As discussed previously, this theory does not contain any gauge-invariant dressed monopole operators as these only contribute at very high order, and hence one only needs to consider the trivial monopole sector. This greatly simplifies the problem and allows for exact computations of the $\mathcal Q$ cohomologies order-by-order.

We will extract monotone and fortuitous cohomologies order-by-order, leaving for the future to write down an index that only counts multi-graviton cohomologies. We  find the first fortuitous BPS cohomologies at order $x^3$. We construct these, and all the other cohomologies up to this order explicitly using the BPS letters. In the following sections, we will first list the single-trace BPS cohomologies, followed by the multi-trace graviton cohomologies, and then finally present the fortuitous BPS cohomologies at order $x^3$.

\subsection{Generalities at large $k$}

The superconformal index \eqref{eq:general_superconf_index} specialised to this case becomes the following four-dimensional integral:
\begin{equation}
    \begin{split}
        \mathcal I_{N=2,k\gg1}(x;u,v) = \frac14\oint&\prod_{i=1}^2\frac{{\rm d}z_i}{2\pi iz_i} \oint\prod_{j=1}^2\frac{{\rm d}w_j}{2\pi iw_j} \\
        &\times \left(1-\frac{z_1}{z_2}\right)\left(1-\frac{z_2}{z_1}\right) \left(1-\frac{w_1}{w_2}\right)\left(1-\frac{w_2}{w_1}\right) \\[1mm]
        &\times \prod_{i,j=1}^2 \mathcal I_{\rm chir}\big(x;u^{\pm1}z_i/w_j,0\big) I_{\rm chir}\big(x;v^{\pm1}w_j/z_i,0\big) ~.
    \end{split}
\end{equation}
As before, there is a decoupled $U(1)$ that one can just trivially integrate over. Since we are going to present the explicit operators in the following sections, we first present the index expanded up to the appropriate order:
\begin{equation}\label{eq:U2_largek_index}
    \begin{split}
        \mathcal I_{N=2,k\gg1}(x;u,v) = 1 &+ \chiT{2}(u)\chiT{2}(v) \, x \\
        &+ \big( 2\chiT{3}(u)\chiT{3}(v) - \chiT{3}(u) - \chiT{3}(v) + \underline{1} - \underline{1} \big) \, x^2 \\
        &+ \big( 2\chiT{4}(u)\chiT{4}(v) - \chiT{4}(u)\chiT{2}(v) - \chiT{2}(u)\chiT{4}(v) \\
        &\qquad - 2\chiT{2}(u)\chiT{2}(v) + \underline{\chiT{4}(u)\chiT{2}(v)} - \underline{\chiT{4}(u)\chiT{2}(v)} \\
        &\qquad + \, \underline{\chiT{2}(u)\chiT{4}(v)} - \underline{\chiT{2}(u)\chiT{4}(v)} + \\
        &\qquad + \, 3\underline{\chiT{2}(u)\chiT{2}(v)} - 3\underline{\chiT{2}(u)\chiT{2}(v)} \big) \, x^3 + \mathcal O(x^4) ~.
    \end{split}
\end{equation}
It is immediately clear that there is a significant cancellation of bosonic and fermionic terms at order $x^3$, and we have included the representations that will appear in our explicit counting.

Let us now discuss the BPS letters and some general comments regarding the computation of the BPS cohomologies. The BPS letters transform in the same way under the $SU(2)_u\times SU(2)_v$ flavour symmetry as in the last section, however they are now $2\times2$ matrices and we will need to assemble them in gauge-invariant operators.

There are two primary ways to find cohomologies:
\begin{enumerate}
\item Using the action of the supercharges that commute with $\mathcal Q$ on existing cohomologies to produce operators at subsequent orders of $x$. This procedure generates the rest of the supermultiplets, starting from some lowest component.

\item Using the $SU(2)_u\times SU(2)_v$ flavour symmetry to constrain the form of the operators and checking for $\mathcal Q$ closedness explicitly.
    
\end{enumerate}
 The advantage of the first approach is that the application of the commuting supercharges automatically produces $\mathcal Q$ closed operators, however linear dependence and $\mathcal Q$ exactness needs to be checked explicitly. On the other hand, the second approach avoids checking for linear dependencies, but $\mathcal{Q}$-closedness must be computed explicitly, which can quickly become cumbersome. We will  present the latter, which is trackable for the first few orders in $x$, but have also checked for consistency by comparing with the first approach.

For later convenience, we set up the notation where the trace over the gauge groups is represented by simple braces, as in the previous section:
\begin{equation}
    (M) \, \coloneq \, {\rm Tr} M ~,
\end{equation}
where $M$ can be a complicated operator made out of the BPS letters such that it has an equal number of bifundamental and anti-fundamental letters. We also define the rank-two and rank-three tensors that implement the symmetrisation of two and three indices, respectively
\begin{equation}
    \begin{split}
        s_{ij}^{pq} &\coloneq \frac12\left( \delta_i^p\delta_j^q + \delta_i^q\delta_j^p \right) ~, \\
        s_{ijk}^{pqr} &\coloneq \frac16\left( \delta_i^p\delta_j^q\delta_k^r + \delta_i^q\delta_j^r\delta_k^p + \delta_i^r\delta_j^p\delta_k^q + \delta_i^r\delta_j^q\delta_k^q + \delta_i^q\delta_j^p\delta_k^r + \delta_i^p\delta_j^r\delta_k^q \right) ~.
    \end{split}
\end{equation}

\subsection{Single-trace cohomologies}\label{sec:st_cohomologies}

We now discuss the single-trace cohomologies in the $U(2)_k\times U(2)_{-k}$ theory. Gauge invariance requires that every bi-fundamental is combined with an anti-bi-fundamental  so we can take as  basic blocks to construct gauge invariants the $2\times2$ matrices obtained by taking their products, for example $AB$, which still transform in the adjoint of the first $U(2)$. Similarly $BA$ is a $2\times 2$ matrix that transforms in the adjoint of the second gauge group. The gauge invariants are traces of products of such building blocks. We will focus on operators with three or less of these basic blocks, as any single-trace operator with more than three basic building blocks can always be decomposed as a sums of multi-trace operators due to trace relations. Also note that a single-trace operator made from a product of three building block, when suitably symmetrised, can also be decomposed via a trace relation. We summarize these trace relations in Appendix \ref{app:tracerelations}.

It is also important to remember that since we are working in the $\mathcal Q$ cohomology, all the operators that we will write down explicitly in what follows are just representatives of the various cohomology classes.
\newline

\noindent{\bf Order $x$} : This order contains only 4 operators, and these are the usual gauge invariants made out of the bosonic BPS letters,
\begin{equation}\label{eq:Largek_AB}
    \SUSU{2}{2} : \, (A_iB_j) ~.
\end{equation}
The difference here compared to the $N=1$ case is that these operators are truly a trace, as the product $A_iB_j$ is itself a $2\times2$ matrix.
\newline

\vspace{1.5em}

\noindent{\bf Order $x^2$} : There are a total of 19 bosons and 7 fermions at this order, as can be seen from \eqref{eq:U2_largek_index}. Only nine of these bosons are single-traces, while all seven of the fermions are single-trace. The bosons are simply the product of four bosonic BPS letters with symmetrisation of the respective $SU(2)$ indices,
\begin{equation}\label{eq:Largek_ABAB}
    \SUSU{3}{3} :\, s_{kl}^{ij}s_{rs}^{pq}(A_iB_pA_jB_q) ~.
\end{equation}
Na\"ively, one might wonder about the absence of $(ABAB)$ single-trace operators in the representations $\SUSU{3}{1}, \SUSU{1}{3}$, and $\SUSU{1}{1}$, where the $\mbf1$ corresponds to an anti-symmetrisation of the indices. However, all three of these can be shown to be $\mathcal Q$ exact justifying their absence in the cohomological counting performed by the index. Indeed, using \eqref{eq:Qaction_fermions}, they can be written as linear combinations of $\mathcal Q (A_i\psi^A_j)$ and $\mathcal Q (B_p\psi^B_q)$.

The seven single-trace fermions correspond to seven fermionic operators of the form $(A\psiA), (B\psiB)$ belonging to the current multiplets of the $SU(2)_u\times SU(2)_v\times U(1)_t$ flavour symmetry. 
The $\SUSU{3}{1}$ and $\SUSU{1}{3}$ in the index are directly identified the $SU(2)_u\times SU(2)_v$  currents
\begin{equation}\label{eq:Largek_XPsi1}
    \SUSU{3}{1} :\, s^{ij}_{kl}\left(A_i\psiA_j\right) \;,\; \SUSU{1}{3} :\, s_{rs}^{pq}\left(B_p\psiB_q\right) ~.
\end{equation}
Meanwhile the final $\SUSU{1}{1}$ corresponds to the $U(1)_t$ current and it the combination 
\begin{equation}\label{eq:Largek_XPsi2}
    \SUSU{1}{1} :\, \epsilon^{ij}\left(A_i\psiA_j-B_i\psiB_j\right) ~.
\end{equation}
The other $\SUSU{1}{1}$ combination
\begin{equation}\label{eq:nonBPS_current}
    \psi_{\rm Non-BPS} \coloneq \epsilon^{ij}\left(A_i\psiA_j+B_i\psiB_j\right)
\end{equation}
fails to be BPS since 

\begin{equation}
    \mathcal Q \psi_{\rm Non-BPS} =\frac{8\pi i}{k} \epsilon_{ij}\epsilon_{rs}(A_iB_rA_jB_s)
\end{equation}
and will play a crucial role in the BPS cohomologies at order $x^3$. It will play a similar role to the one played by the Konishi multiplet in $\mathcal{N}=4$ SYM \cite{Chang:2022mjp,Choi:2023znd,Budzik:2023vtr}. Note that this current is in fact BPS in the $N=1$ ABJM theory, because the right-hand side vanishes, and its non BPS-ness in the $N>1$ theory (at $k>1$) justifies the interpretation of the corresponding $N=1$ operator as being fortuitous.

\vspace{1.5em}

\noindent{\bf Order $x^3$} : The most natural single-trace object to consider at this order would be a product of the three $A$ and three $B$. However, this decomposes into a linear combination of multi-trace operators due to a simple trace relation, as we will see explicitly in the next section.

Therefore, the only bosonic single-trace operators at this order can either be $D(AB)$, $(ADB)$, or $(\psiA\psiB)$. By an explicit computation, it can be shown that the $\mathcal Q$ action on $\big(BDA\big)$ gives
\begin{equation}
    \mathcal Q \big(B_pDA_i\big) = i\mathcal Q\big(\psiA_i\psiB_p\big) ~.
\end{equation}
Therefore we conclude that the two sets of bosonic single-trace operators are
\begin{equation}\label{eq:Largek_DAB}
    \SUSU{2}{2} :\, D\big(A_iB_p\big) \;,\; \SUSU{2}{2} :\, \big(iB_pDA_i + \psiA_i\psiB_p\big) ~.
\end{equation}
This particular combination can also be confirmed by using the action of the supercharges that commute with $\mathcal Q$ on the fermionic currents at order $x^2$.

We now turn to the fermionic operators. They are of the form $(\psiA ABA)$ and $\psiB BAB$, and we can group them in $SU(2)_u\times SU(2)_v$ representations as follows:
\begin{equation}\label{eq:SingleTraceFermions_x3}
    \begin{split}
        \SUSU{4}{2} :\, s_{lmn}^{ijk}\big(\psiA_iA_jB_pA_k\big) \;,\; \SUSU{2}{2} :\, \epsilon^{jk}\big(\psiA_iA_jB_pA_k\big) \;,\; \SUSU{2}{2} :\, \epsilon^{ij}\big(\psiA_iA_jB_pA_k\big) ~, \\
        \SUSU{2}{4} :\, s_{stu}^{pqr}\big(\psiB_pB_qA_iB_r\big) \;,\; \SUSU{2}{2} :\, \epsilon^{qr}\big(\psiB_pB_qA_iB_r\big) \;,\; \SUSU{2}{2} :\, \epsilon^{pq}\big(\psiB_pB_qA_iB_r\big) ~.
    \end{split}
\end{equation}
We now need to determine the combinations of these that are $\mathcal Q$ closed. From representation theory, the operators in $\SUSU{4}{2}$ and $\SUSU{2}{4}$ cannot recombine with anything at this order and can be further checked explicitly to be $\mathcal Q$ closed. The four different sets of operators in the $\SUSU{2}{2}$ representations can all mix, and in fact only two of them are $\mathcal Q$ closed. Furthermore, one of them can be shown to be $\mathcal Q$ exact,
\begin{equation}
    \epsilon^{ij}\big(\psiA_pA_iB_qA_j + \psiB_qB_iA_pB_j\big) = \frac{k}{4\pi i} \mathcal Q\big(\psiA_p\psiB_q\big) ~.
\end{equation}
Thus we only have  a single $\SUSU{2}{2}$ operator fermionic single-trace at order $x^3$
\begin{equation}\label{eq:Largek_PsiXXX}
    \SUSU{2}{2} :\, \epsilon^{ij}\big(\psiA_iA_jB_qA_p - \psiB_iB_jA_pB_q + \psiA_pA_iB_qA_j\big) ~.
\end{equation}

We summarise these operators with their representations in Table \ref{table2} at the end of the next section along with the contributions from the multi-trace operators that we consider next.

\subsection{Multi-trace cohomologies}

The single-trace operators counted in the previous section clearly describe only a part of the BPS spectrum visible in the superconformal index \eqref{eq:U2_largek_index}. In this section, we will describe the multi-trace BPS operators that match further contributions to the superconformal index. The operators in the previous section and those that we will discuss in this section are all BPS without the use of a finite $N$ trace relations, and in fact remain BPS in the corresponding $N\rightarrow\infty$ theory. We will reconsider this in the next section to identify the fortuitous states in the $N=2$ large $k$ theory by comparing the contributions in the previous and present section to the superconformal index \eqref{eq:U2_largek_index}.

Although we do not consider operators that are $\mathcal{Q}$-closed due to a finite $N$ trace relation (these would be fortuitous operators), the trace relations at $N=2$ still play an important role in the counting of multi-gravitons at order $x^3$. This is because multi-trace operators can be $\mathcal{Q}$-exact due to a finite $N$ trace relations, and we need to take this into account to obtain the correct count of cohomologies.
\newline

\noindent{\bf Order $x$} : This is trivial since the only operators allowed at this order are single-traces and they have already been discussed in the previous section. There is nothing to add here.
\newline

\noindent{\bf Order $x^2$} : Through the single-trace operators, we have already accounted for the bosonic $\SUSU{3}{3}$, and all the fermionic contributions to the index \eqref{eq:U2_largek_index}. Thus we have a remaining $\SUSU{3}{3}$ and $\SUSU{1}{1}$ representations which are both multi-trace operators. These are quite natural and correspond to
\begin{equation}\label{eq:Largek_ABAB_multi}
    \SUSU{3}{3} :\, s_{kl}^{ij}s_{rs}^{pq}\big(A_iB_p\big)\big(A_jB_q\big) \;,\; \SUSU{1}{1} :\, \epsilon^{ij}\epsilon^{pq}\big(A_iB_p\big)\big(A_jB_q\big) ~.
\end{equation}
Note that while the corresponding $\SUSU{1}{1}$ single-trace operator was $\mathcal Q$ exact, the double-trace nature of this $\SUSU{1}{1}$ prevents such a relation. Additionally, in this case the double-trace $\SUSU{3}{1}$ and $\SUSU{1}{3}$ which came from fermionic operators cannot be made into double-traces in a gauge-invariant way.
\newline

\noindent{\bf Order $x^3$} : This is the most interesting order, in part, due to the poor match by the single-trace operators. We first check that the single-trace operators of the form: \begin{equation}
    (A_iB_jA_kB_lA_mB_n) ~,
\end{equation}
are actually decomposable as products of smaller traces. We can assume that the indices $(ikm)$ and $(jln)$ are fully symmetrized due to the following relation (and a similar one with $A$ and $B$ interchanged)
\begin{equation}\label{eq:Superpot_relation}
    A_iB_1A_jB_2 - A_iB_2A_jB_1 = \frac{k}{2\pi i}\mathcal Q \, A_i\psiA_j ~,
\end{equation}
implying that the the two terms of the left hand side are identified in the $\mathcal Q$-cohomology. Furthermore, the symmetrised product of three $(AB)$ blocks in a single-trace decomposes into multi-trace operators due to a trace relation at $N=2$ in the following way (see also Appendix \ref{app:tracerelations}): 
\begin{equation}\label{eq:TraceRelation1}
    \begin{split}
        (A_iB_jA_iB_jA_pB_q) = (A_iB_j)(A_iB_jA_pB_q) &+ \frac12(A_pB_q)(A_iB_jA_iB_j) \\
        &- \frac12(A_iB_j)(A_iB_j)(A_pB_q) ~.
    \end{split}
\end{equation}
From this, it is clear that all these operators reduce to double and triple-trace operators made out of the 6 bosonic BPS letters, and hence we only need to count the remaining multi-trace operators.

There are a total of 4 operators of type $(AB)$ at order $x$ and 9 operators of type $(AB)(AB)$ at order $x^2$. The bosonic multi-trace operators at order $x^3$ are formed by the symmetrised product of these operators. Thus, the total number of these multi-trace operators are
\begin{equation}
    \begin{split}
        20\times(AB)(AB)(AB) \;,\; 36\times (AB)(ABAB) ~.
    \end{split}
\end{equation}
These 56 operators are, however, not all independent in the $\mathcal Q$ cohomology, and in fact, there are four bosonic relations between them.
These relations are due to the following expression being $\mathcal Q$ exact:\
\begin{equation}\label{eq:4bosonic_rel}
    \epsilon^{rs}\epsilon^{pq}\big(A_iB_jA_rB_pA_sB_q\big) = \frac{4\pi i}{k}\,\epsilon^{rs}\mathcal Q\big(A_iB_jA_r\psiA_s\big) ~,
\end{equation}
where it is understood that the left hand side can be decomposed into a linear combination of multi-trace operators by using \eqref{eq:Superpot_relation} and \eqref{eq:TraceRelation1}. The right hand side of \eqref{eq:4bosonic_rel} clearly transforms in the $\SUSU{2}{2}$ representation of the $SU(2)_u\times SU(2)_v$ flavour symmetry, and hence we see that a combination of the multi-trace operators here satisfy a relation in the $\mathcal Q$ cohomology.
For completeness, we can now list the 52 bosonic multi-trace operators that contribute to the superconformal index at order $x^3$. The triple-trace operators can be organised as
\begin{equation}\label{eq:Largek_ABABAB_triple}
    \SUSU{4}{4} :\, s^{ijk}_{lmn}s^{pqr}_{stu}(A_iB_p)(A_jB_q)(A_kB_r) \;,\; \SUSU{2}{2} :\, \epsilon^{ij}\epsilon^{pq}(A_iB_p)(A_jB_q)(A_kB_r) ~.
\end{equation}
While the double-trace operators can be organised as
\begin{equation}\label{eq:Largek_ABABAB_double}
    \begin{split}
        \SUSU{4}{4} &:\, s^{ijk}_{lmn}s^{pqr}_{stu}(A_iB_p)(A_jB_qA_kB_r) \\
        \SUSU{4}{2} &:\, s^{ijk}_{lmn}\epsilon^{qr}(A_iB_pA_jB_q)(A_kB_r) \;,\; \SUSU{2}{4} :\, s^{pqr}_{stu}\epsilon^{jk}(A_iB_pA_jB_q)(A_kB_r) ~,
    \end{split}
\end{equation}
totalling to 52 bosonic multi-trace operators, exactly as expected.\footnote{This results exactly matches the counting that the corresponding Hilbert series of the moduli space of $U(2)_k\times U(2)_{-k}$ ABJM theory at large $k$ produces \cite{Cremonesi:2016nbo}.}

Finally, the multi-trace fermionic operators are easy to list as these just correspond to the product between the seven fermionic BPS currents at order $x^2$ times the meson at order $x$.
\begin{equation}\label{eq:Largek_PsiXXX_1}
    \SUSU{2}{2} :\, \epsilon^{ij}\big(A_i\psiA_j-B_i\psiB_j\big)\big(A_kB_l\big)
\end{equation}
The product of the $\SUSU{3}{1}$ triplet fermionic current with the mesons decomposes at the level of representation theory as $\SUSU{3}{1}\times\SUSU{2}{2} = \SUSU{4}{2}+\SUSU{2}{2}$, and similarly for the $\SUSU{1}{3}$ triplet fermionic current. Therefore the remaining multi-trace operators of this form can be expressed as
\begin{equation}\label{eq:Largek_PsiXXX_2}
    \begin{split}
   \SUSU{2}{2} :\, \epsilon^{ij}\big(A_i\psiA_j-B_i\psiB_j\big)&\big(A_kB_l\big) ~ , \qquad \, \,\\
        \SUSU{4}{2} :\, s^{ijk}_{lmn}\big(\psiA_iA_j\big)\big(B_pA_k\big) \;&,\; \SUSU{2}{4} :\, s^{pqr}_{stu}\big(\psiB_pB_q\big)\big(A_iB_r\big) ~, \\
        \SUSU{2}{2} :\, s^{ij}_{qr}\epsilon^{ik}\big(\psiA_iA_j\big)\big(A_kB_p\big) \;&,\; \SUSU{2}{2} :\, s^{pq}_{jk}\epsilon^{pr}\big(\psiB_pB_q\big)\big(B_rA_i\big) ~.
    \end{split}
\end{equation}
All these operators contribute to the index since there is no relation between these fermionic operators and none of them are $\mathcal Q$ exact.

Let us take summarise our cohomological counting of operators in the $U(2)_k\times U(2)_{-k}$ ABJM theory at large $k$ and compare it to the superconformal index. The operators we have counted so far correspond to either BPS single-traces themselves or their products. Therefore, these constitute the multi-graviton spectrum of the $N=2$ theory up to order $x^3$. We schematically summarise the multi-graviton cohomologies with their contributions to be index in Table \ref{table2}.
\begin{table}[!h]
    \begin{center}
        \begin{tabular}{|c|c|c|c|c|} 
            \hline
            $x^{\mathcal D+J_3}$ & Operator & $(-1)^F$\,Total & $SU(2)_u\times SU(2)_v$ rep. & Reference \\
            \hline\hline
            $x$ & $(AB)$ & 4 & $\SUSU{2}{2}$ & \eqref{eq:Largek_AB} \\
            \hline\hline
            \multirow{3}{*}{$x^2$} & $(ABAB)$ & 9 & $\SUSU{3}{3}$ & \eqref{eq:Largek_ABAB} \\
            & $(AB)(AB)$ & 10 & $\SUSU{3}{3}+\SUSU{1}{1}$ & \eqref{eq:Largek_ABAB_multi} \\
            & $(\psiA A)$ & $-7$ & $\SUSU{3}{1}+\SUSU{1}{3}+\SUSU{1}{1}$ & \eqref{eq:Largek_XPsi1}+\eqref{eq:Largek_XPsi2}  \\
            \hline\hline
            \multirow{6}{*}{$x^3$} & $(ABAB)(AB)$ & 32 & $\SUSU{4}{4}+\SUSU{4}{2}+\SUSU{2}{4}$ & \eqref{eq:Largek_ABABAB_double} \\
            & $(AB)(AB)(AB)$ & 20 & $\SUSU{4}{4}+\SUSU{2}{2}$ & \eqref{eq:Largek_ABABAB_triple} \\
            & $D(AB)$ & 4 & $\SUSU{2}{2}$ & \eqref{eq:Largek_DAB} \\
            & $(ADB+\psiA\psiB)$ & 4 & $\SUSU{2}{2}$ & \eqref{eq:Largek_DAB} \\
            & $(\psi^A ABA)$ & $-20$ & $\SUSU{4}{2}+\SUSU{2}{4}+\SUSU{2}{2}$ & \eqref{eq:SingleTraceFermions_x3}+\eqref{eq:Largek_PsiXXX} \\
            & $(\psiA A)(AB)$ & $-28$ & $\SUSU{4}{2}+\SUSU{2}{4}+3\times\SUSU{2}{2}$ & \eqref{eq:Largek_PsiXXX_2} \\
            \hline
        \end{tabular}
    \end{center}
    \caption{Multi-graviton cohomologies in the $U(2)_k\times U(2)_{-k}$ ABJM theory at large $k$. The operator column is schematic and one should consider the possible swaps of $A\leftrightarrow B$ when applicable.}\label{table2}
\end{table}
Using this, we determine the multi-graviton index of the this theory to be
\begin{equation}
    \begin{split}
        \mathcal I_{N=2,k\gg1}^{\rm multi-grav}(x;u,v) = 1 &+ \chiT{2}(u)\chiT{2}(v) \, x \\
        &+ \big( 2\chiT{3}(u)\chiT{3}(v) - \chiT{3}(u) - \chiT{3}(v) + \underline{1} - \underline{1} \big) \, x^2 \\
        &+ \big( 2\chiT{4}(u)\chiT{4}(v) - \chiT{4}(u)\chiT{2}(v) - \chiT{2}(u)\chiT{4}(v) \\
        &\qquad - \chiT{2}(u)\chiT{2}(v)\big)x^3+ \mathcal O(x^4) ~.
    \end{split}
\end{equation}
Upon comparing this to the full superconformal index \eqref{eq:U2_largek_index}, we notice a mismatch at order $x^3$,
\begin{equation}
    \mathcal I_{N=2,k\gg1}(x;u,v) - \mathcal I_{N=2,k\gg1}^{\rm multi-grav}(x;u,v) = - \chiT{2}(u)\chiT{2}(v) x^3 + \mathcal O(x^4) ~.
\end{equation}
These are precisely the fortuitous states in the $N=2$ large $k$ ABJM theory that we discuss next.

\subsection{Fortuitous cohomologies}

The mismatch between the superconformal index and the multi-graviton index of the $N=2$ large $k$ ABJM theory at order $x^3$ is a clear indication of fortuity. However, we should be able to construct these states and exhibit the property that their BPS-ness follows from a special $N=2$ trace relation. We do this next and present the lowest lying fortuitous operator in this ABJM theory.

The recipe to construct the fortuitous operator is to consider the product of the non-BPS current \eqref{eq:nonBPS_current} and the mesons $(AB)$. Although this transforms in the correct representation: $\SUSU{2}{2}$, it can be easily checked that this is not $\mathcal Q$-closed, and hence not BPS. It turns out that combining this with a specific combination of the non $\mathcal Q$-closed operators in \eqref{eq:SingleTraceFermions_x3} produces a $\mathcal Q$-closed operator,
\begin{equation}
    \mathcal O_f = \epsilon^{ij}\left(\psiA_iA_jB_kA_l+\psiB_iB_jA_lB_k\right)-\frac{3}{4}\,\epsilon^{ij}\left(B_kA_l\right)\left(\psiA_iA_j+\psiB_iB_j\right) ~.
\end{equation}
We can now explicitly check that $\mathcal O_f$ is $\mathcal Q$ closed due to a trace relation for $2\times2$ matrices. We have
\bea
    \mathcal{Q}\,\mathcal O_f &=& \epsilon^{pq}\epsilon^{ij}\left(B_pA_iB_qA_jB_kA_l-A_pB_iA_qB_jA_lB_k\right) \notag \\
    &-&\frac{3}{4}\,\epsilon^{pq}\epsilon^{ij}\left(B_kA_l\right)\left(B_pA_iB_qA_j-A_pB_iA_qB_j\right) ~.
\eea
We now use \eqref{tracerelation}, which in this case reads
\bea \label{tracerelationsec4}
(B_p A_i B_q A_j B_k A_l )+( B_q A_j B_p A_i B_k A_l ) &=&  (B_p A_i)(B_q A_j B_k A_l)+( B_q A_j)(B_p A_i B_k A_l) \notag \\
&+&(B_k A_l)(B_p A_i B_q A_j)-(B_p A_i)(B_q A_j)( B_k A_l) \notag \\
\eea
Applying the trace relations to all the terms, we get that $\mathcal{Q}\mathcal O_f$ vanishes.

Recall that \eqref{tracerelationsec4} holds only for $2\times2$ matrices, and hence this operator will no longer be BPS at $N>2$, as expected from fortuitous states. Moreover, based on the one-loop $\mathcal Q$ action on the BPS letters \eqref{eq:Qaction_fermions}, it is evident that $\mathcal O_f$ is not $\mathcal Q$-exact. Thus, this is a genuine BPS operator in the $\SUSU{2}{2}$ representation by construction, and represents the missing $\SUSU{2}{2}$ cohomology at order $x^3$ that fully reproduces the superconformal index up to this order.

As remarked earlier, the fortuitous operator $\mathcal O_f$ is built out of two non-BPS operators such that this specific linear combination becomes BPS at $N=2$. Moreover, one of these components is a single-trace operator, in contrast to the fortuitous state found in 4d $\mathcal N=4$ super Yang--Mills \cite{Chang:2022mjp,Choi:2023znd,Budzik:2023vtr}, where a triple-trace operator is the representative of the lowest lying fortuitous cohomology. It does share a similar feature with the operator found there: it is a non-BPS operator dressed by supergravitons. 

Note that the crucial non-BPS piece in $\mathcal O_f$ is the non-BPS fermionic current $\psi_{\rm non-BPS}$. This is a singlet and it multiples a meson to yield a $\SUSU{2}{2}$ representation -- once as a single-trace and once as a double-trace.

\section{Fortuity at strong coupling: $U(2)_2 \times U(2)_{-2}$
\label{sec:5}}

In this section, we explore fortuity in the strongly coupled regime, focusing on $U(2)_2\times U(2)_{-2}$ ABJM theory. This theory contains monopole operators and it has a supersymmetry enhancement to $\mathcal N=8$. This supersymmetry enhancement is precisely caused by dressed monopole operators, but the presence of monopole operators is what makes the overall analysis challenging. Supersymmetry enhancement will be a very valuable guide in what follows.

The main tool at our disposal will be the $SU(4)$ flavour symmetry that the superconformal index sees in this case. It is essential to note that the definition of the supersymmetry action and the notion of single- and multi-traceness is quite subtle for dressed monopole operators, making it challenging to define fortuitous operators. Nevertheless, we provide a self-consistent count of the BPS cohomologies visible in the superconformal index by relying on the $SU(4)$ symmetry. With our definition of graviton cohomologies, this lets us identify fortuitous cohomologies in this theory. By restricting to the zero monopole sector and using the branching rules from $SU(4)$ to $SU(2)_u\times SU(2)_v\times U(1)_t$, we recover the fortuitous cohomologies in the weakly coupled theory, providing a non-trivial consistency check of our cohomological counting.

\subsection{Superconformal index and monopole operators}

The superconformal index \eqref{eq:general_superconf_index} specialised to this case becomes the following four-dimensional integral:
\begin{equation}
    \begin{split}
        \mathcal I_{N=2,k=2}(x;u,v,t=t_1t_2) = &\sum_{(\mbf m,\widetilde{\mbf m})\in\mathbb Z^4}\frac14\oint\prod_{i=1}^2\frac{{\rm d}z_i}{2\pi iz_i} z_i^{2m_i} \oint\prod_{j=1}^2\frac{{\rm d}w_j}{2\pi iw_j} w_i^{2\widetilde m_i} \\
        &\quad \times \left(1-\frac{z_1}{z_2}\right)\left(1-\frac{z_2}{z_1}\right) \left(1-\frac{w_1}{w_2}\right)\left(1-\frac{w_2}{w_1}\right) \\[1mm]
        &\quad \times \; t_1^{\sum_i m_i}t_2^{\sum_j\widetilde m_j} \prod_{i,j=1}^2 \mathcal I_{\rm chir}\big(x;u^{\pm1}z_i/w_j,m_i-\widetilde m_j\big) \\
        &\qquad\qquad\quad \times\; \mathcal I_{\rm chir}\big(x;v^{\pm1}w_j/z_i,m_i-\widetilde m_j\big) ~.
    \end{split}
\end{equation}
As already explained in Section \ref{subsec:superconf_index}, integrating over the decoupled diagonal $U(1)$ leads to the constraint $m_1+m_2=\widetilde m_1+\widetilde m_2$. The superconformal index expanded to the third order can be expressed as
\begin{equation}\label{eq:N2k2_index}
    \begin{split}
        \mathcal I_{N=2,k=2}(x;a_i) = 1 &+ \chiSU{10} \,x + \Big( 2\chiSU{35} + \chi_{\mbf{20}^{'}}^{SU(4)} - \chi_{\mbf{15}}^{SU(4)} \Big)\,x^2 \\
        &+ \Big( 2\chiSU{84^{''}} + \chiSU{140^{''}} + \chiSU{126} + \underline{\chiSU{10}} + \underline{\chiSU{6}} \\
        &\qquad\quad - 2\chiSU{70} - \chi_{\mbf{64}}^{SU(4)} - \chiSU{10} - \underline{\chiSU{10}} - \underline{\chiSU{6}} \Big) \,x^3 \\
        &+ \mathcal O(x^4) ~.
    \end{split}
\end{equation}
We will follow the same fugacity map between $SU(4)_{a_i}$ and $SU(2)_u\times SU(2)_v\times U(1)_t$ that we used in Section \ref{sec:3}, and recall it for convenience
\begin{equation}
    a_1 = \frac{u}{t^{\frac12}} \;,\; a_2 = \frac{1}{u\,t^{\frac12}} \;,\; a_3 = v\,t^{\frac12} \;,\; a_4 = \frac{t^{\frac12}}{v} ~.
\end{equation}
As we will see, the only monopole fluxes that contribute up to order $x^3$ are those with $m_i=\widetilde {m}_i$, and thus the bare monopoles contributing up to this order have vanishing $R$-charge due to formula \eqref{Rchargemonopole}.

We will take inspiration from an approach that proved successful in the study of the Hilbert series \cite{Cremonesi:2013lqa,Cremonesi:2016nbo}. A magnetic flux configuration breaks the gauge group down to the subgroup $H_{(\mbf m,\widetilde{\mbf m})}$ that commutes with the fluxes. The monopole carries electric charge under the abelian part of the residual gauge group given by the embedding of  $(-k\,\mbf m,k\,\widetilde{\mbf m})$ into the center of $H_{(\mbf m,\widetilde{\mbf m})}$. The fields used to dress the bare monopole are the massless fields in this flux background. They must transform under the abelian part of the residual gauge group to compensate for the electric charge of the monopole and are otherwise gauge invariant. Thus, the dressed operators can be viewed as local operators living in the residual gauge theory defined by a fixed monopole background specified by the magnetic fluxes $(\mbf m,\widetilde{\mbf m})$.

Concretely in our case, for magnetic fluxes with $m_i=\widetilde {m}_i$ and of the form $m_1=m_2$, the residual gauge symmetry algebra is $U(2)\times U(2)/U(1)_{\rm dec}\cong SU(2)\times SU(2)\times U(1)$. On the other hand for $m_1 \neq m_2$, the residual gauge theory is simply $U(1)^2\times U(1)^2$. It is evident from this discussion that the definition of a multi-trace operator is not clear in the presence of certain non-trivial monopole backgrounds, in particular for $m_1\neq m_2$. We will identify such operators as single-trace or multi-trace by using $SU(4)$ representation theory as we will explain in the next section. Purely bosonic BPS operators made with the letters $Y$ and no derivatives correspond to the chiral ring of the ABJM theory.  Their number has been already counted in \cite{Cremonesi:2016nbo} and this will serve as a non-trivial check of our results.

\begin{table}[!h]
    \begin{center}
        \begin{tabular}{|c|c|c|} 
            \hline
            Dressing fields & $\left(U(1)\times U(1)\right)^{(1)}$ & $\left(U(1)\times U(1)\right)^{(2)}$ \\
            \hline\hline
            $A_{1,2}^{(1)}$ & $(-1,0)$ & $(0,0)$ \\
            \hline
            $B_{1,2}^{(1)}$ & $(1,0)$ & $(0,0)$ \\
            \hline
            $A_{1,2}^{(2)}$ & $(0,0)$ & $(0,-1)$ \\
            \hline
            $B_{1,2}^{(2)}$ & $(0,0)$ & $(0,1)$ \\
            \hline\hline
            $\psiA_{1,2}{}^{(1)}$ & $(1,0)$ & $(0,0)$ \\
            \hline
            $\psiB_{1,2}{}^{(1)}$ & $(-1,0)$ & $(0,0)$ \\
            \hline
            $\psiA_{1,2}{}^{(2)}$ & $(0,0)$ & $(0,1)$ \\
            \hline
            $\psiB_{1,2}{}^{(2)}$ & $(0,0)$ & $(0,-1)$ \\
            \hline
        \end{tabular}
    \end{center}
    \caption{Gauge charges of the various massless fields in the $U(1)^2\times U(1)^2$ residual gauge theory in the monopole background labeled by magnetic fluxes $m_1\neq m_2$.}\label{table_residualcharges}
\end{table}

\subsection{Graviton cohomologies}

Considering magnetic fluxes with $m_1\neq m_2$, the residual gauge theory contains two $U(1)\times U(1)$ quivers, that we disintguish by another label $(1)$ or $(2)$. In this case, the massless fields in the monopole background used to dress the bare monopoles also carry these labels. These two $U(1)\times U(1)$ quivers are paired such that the gauge charges of the residual fields are as presented in Table \ref{table_residualcharges}. At a more schematic level, the four bosonic, $Y$, and fermionic, $\Psi$, BPS letters transform as the $\overline{\mbf 4}$ and $\mbf 4$ representations of the $SU(4)$ flavour symmetry, respectively. Let us now go order-by-order as before, and present the cohomological counting of the operators that contribute to the superconformal index. \vspace{1em}

\noindent{\bf Order $x$} : At this order, the 10 operators can be  schematically written as 
\begin{equation}\label{eq:N2k2_YY}
    \overline{\mbf{10}} :\, \big(YY\big) ~.
\end{equation}
The monopole sectors contributing to this order\footnote{Magnetic fluxes are defined up to the Weyl group so $(1,0)$ is gauge equivalent to $(0,1)$.} are $\mbf m=(0,0)$, $(1,0)$, and $(-1,0) $ corresponding to the following decomposition of the $\overline{\mbf{10}}$ representation of $SU(4)$ into $SU(2)_u\times SU(2)_v\times U(1)_t$:
\begin{equation}\label{eq:U2_orderx}
    \SUSU{3}{1}_{-1} : \, (T^{(-1,0)}AA)^{(1)} \;,\; \SUSU{2}{2}_0 : \, (AB) \;,\; \SUSU{1}{3}_1 : \, (T^{(1,0)}BB)^{(1)} ~.
\end{equation}
Notice that the antisymmetric combination $\mbf 6$ in the product of two letters $Y$, $\overline{\mbf 4}\times \overline{\mbf 4} = \overline{\mbf{10}}+\mbf 6$, is not present. For $N=1$ this was due to the commutation of the $Y$. Here it follows from the absence of suitable dressed monopole operators for non-zero magnetic flux. As we will see, the same arguments enforce a symmetrization of the $Y$ in the full graviton spectrum, at least at low orders. In this case, the dressed monopole operators can safely be interpreted as single-trace operators. The results at this order are summarised in Table \ref{table_N2k2_orderx}.
\begin{table}[!h]
    \begin{center}
        \begin{tabular}{|c|c|c|c|c|} 
            \hline
            $U(1)_t$ & Operator & $(-1)^F$\,Total & $SU(2)_u\times SU(2)_v$ \\
            \hline\hline
            0 & $(AB)$ & 4 & $\SUSU{2}{2}$ \\
            \hline
            1 & $\big(T^{(1,0)}BB\big)^{(1)}$ & 3 & $\SUSU{1}{3}$ \\
            \hline
            -1 & $\big(T^{(-1,0)}AA\big)^{(1)}$ & 3 & $\SUSU{3}{1}$ \\
            \hline
        \end{tabular}
    \end{center}
    \caption{Cohomology representatives contributing to the $U(2)_2\times U(2)_{-2}$ ABJM theory at order $x$. }
    \label{table_N2k2_orderx}
\end{table}
\vspace{1.5em}

\noindent{\bf Order $x^2$} : At this order, we start encountering both single and multi-trace operators analogously to the corresponding large $k$ theory. We first consider the single-trace bosonic operators of the schematic form $(YYYY)$. Once again, the non-vanishing representation of this kind is the fully symmetric one and corresponds to the $\overline{\mbf{35}}$, which matches a contribution in the index \eqref{eq:N2k2_index}: 
\begin{equation}\label{eq:N2k2_YYYY}
    \overline{\mbf{35}} :\, \big(YYYY\big) ~.
\end{equation}

This order also contains fermionic single-trace operators of the schematic form $(\Psi Y)$ which fit in the $\mbf{15}$ representation of $SU(4)$ following from $\mbf 4\times\overline{\mbf 4}=\mbf{15}+\mbf 1$.
\begin{equation}\label{eq:N2k2_PsiY}
    \mbf{15} :\, \big(\Psi Y\big) ~.
\end{equation}
The singlet here can be explicitly shown to be non-BPS, as it corresponds to an operator in the trivial flux sector and was already dealt with in the previous section.

Let us now discuss the multi-trace operators of the schematic form $(YY)(YY)$. These would correspond to the symmetric combinations of the 10 $(YY)$ from order $x$, and hence we expect $55$ of these.

As manifest $SU(4)$ representations, these 55 operators appear in the decomposition $\overline{\mbf{10}} \times \overline{\mbf{10}} = \overline{\mbf{35}} + \mbf{20^{'}} + \overline{\mbf{45}}$, and therefore $\overline{\mbf{45}}$ should go away and we are left with
\begin{equation}\label{eq:N2k2_YY_YY}
    \overline{\mbf{35}} + \mbf{20^{'}} :\, \big(YY \big)\big(YY \big) \, .
\end{equation}
This agrees with the Hilbert series counting \cite{Cremonesi:2016nbo} which predicts in total $90$ purely bosonic BPS operators.

The various contributions can be grouped together by their topological charges and monopole backgrounds by using the decomposition of the corresponding $SU(4)$ representations, as summarised in Table \ref{table_N2k2_orderx2}. Combining the exact match of our counting with the superconformal index \eqref{eq:N2k2_index}, and the match between the operators made out of purely bosonic BPS letters with the prediction for the chiral ring coming from the Hilbert series \cite{Cremonesi:2016nbo}, serves as a non-trivial consistency check of our cohomological counting.

\begin{table}[!h]
    \centering
        \begin{tabular}{|c|c|c|c|c|} 
            \hline
            $U(1)_t$ & Operator\footnotemark & $(-1)^F$\,Total & $SU(2)_u\times SU(2)_v$ \\
            \hline\hline
            \multirow{2}{*}{$0$} & $(ABAB)+(AB)(AB)$ & $19$ & $\ma{\SUSU{3}{3}}+\bl{\SUSU{3}{3}}+\ma{\SUSU{1}{1}}$ \\ &$(\psi^A A)+(\psi^B B)$ &-7 & $ \ora{-\SUSU{3}{1}-\SUSU{1}{3}-\SUSU{1}{1}}$  \\
            & $\big(T^{(1,-1)}A^{(1)}A^{(1)}B^{(2)}B^{(2)}\big)$ & $9$ & \ma{$\SUSU{3}{3}$} \\
            \hline
            \multirow{2}{*}{$2$} & $\big(T^{(2,0)}BBBB\big)^{(1)}$ & $5$ & \bl{$\SUSU{1}{5}$} \\
            & $\big(T^{(1,1)}{\rm det}B\,{\rm det}B\big)$ & $6$ & $\ma{\SUSU{1}{5}+\SUSU{1}{1}}$ \\
            \hline
            \multirow{3}{*}{$1$} & $\big(T^{(1,0)}BBAB\big)^{(1)}$ & $8$ & $\bl{\SUSU{2}{4}}$ \\
            & $\big(T^{(1,0)}BB\big)^{(1)}(AB)^{(2)}$ & $12$ & $\ma{\SUSU{2}{4}+\SUSU{2}{2}}$ \\
            & $\big(T^{(1,0)}B\psiA\big)^{(1)}$ & $-4$ & $-\ora{\SUSU{2}{2}}$ \\
            \hline
            \multirow{3}{*}{$-1$} & $\big(T^{(-1,0)}AABA\big)^{(1)}$ & $8$ & $\bl{\SUSU{4}{2}}$ \\
            & $\big(T^{(-1,0)}AA\big)^{(1)}(AB)^{(2)}$ & $12$ & $\ma{\SUSU{4}{2}+\SUSU{2}{2}}$ \\
            & $\big(T^{(-1,0)}A\psiB\big)^{(1)}$ & $-4$ & $-\ora{\SUSU{2}{2}}$ \\
            \hline
            \multirow{2}{*}{$-2$} & $\big(T^{(-2,0)}AAAA\big)^{(1)}$ & $5$ & $\bl{\SUSU{5}{1}}$ \\
            & $\big(T^{(-1,-1)}{\rm det}A\,{\rm det}A\big)$ & $6$ & $\ma{\SUSU{5}{1}+\SUSU{1}{1}}$ \\
            \hline
        \end{tabular}
    \caption[Cohomology reps for the U(2)xU(2) ABJM theory at order x2]{Cohomology representatives contributing to the $U(2)_2\times U(2)_{-2}$ ABJM theory at order $x^2$. The representations in blue combine to give the $\overline{\mbf{35}}$ representation of $SU(4)$ corresponding to the single-trace operators $\bl{(YYYY)}$, the ones in magenta give the $\overline{\mbf{35}} + \mbf{20^{'}}$ representation corresponding to the multi-trace operators $\ma{(YY)(YY)}$, and the ones in orange give the ${\mbf 15}$
   representation corresponding to $\ora{(\Psi Y)}$.} \label{table_N2k2_orderx2}
\end{table}

\vspace{1.5em}

\noindent{\bf Order $x^3$} : From the analysis at large $k$ in the previous section, we expect trace relations to start playing a role at this order. This significantly complicates the analysis as our approach is based on exploiting the $SU(4)$ flavour symmetry to avoid a detailed characterisation of the relations between dressed monopole operators. Therefore, we only present a consistent cohomological counting at order $x^3$ without being able to explicitly write down the series of operators and the relations that they satisfy.

From the large $k$ analysis, we expect two kinds of  bosonic single-trace operators at this order: 1) those coming from $D$ acting on the 10 $(YY)$ operators, and 2) those coming from the product of two fermionic BPS letters $(\Psi\Psi)$, which following from the large $k$ analysis are naturally paired with $(YDY)$ in the BPS cohomologies. Therefore we have
\begin{equation}\label{eq:N2k2_DYY}
    \overline{\mbf{10}} :\, D(YY) \;,\; \mbf 6 :\, iYDY+\Psi\Psi ~.
\end{equation}
These representations correctly reproduce the trivial flux sector operators listed in Table \ref{table2}, while the corresponding dressed monopole operators are listed in Table \ref{table_N2k2_orderx3}.

The remaining bosonic multi-trace operators contain double-trace and triple-trace operators of the form $\big(YYYY\big)\big(YY\big)$ and $\big(YY\big)\big(YY\big)\big(YY\big)$, respectively. We expect some relations between these multi-trace operators, as already evident in the zero flux sector, as shown in \eqref{eq:TraceRelation1}. However, determining the exact relations between the dressed monopole operators is quite challenging. Therefore, we will use the chiral ring counting from the Hilbert series \cite{Cremonesi:2016nbo} to determine the number of relations and hence the correct number of multi-trace operators.

\footnotetext[18]{The ${\rm det} A$ are the appropriate dressing operators for the bare monopole $T^{(-1,-1)}$ as these are manifestly invariant under the residual $SU(2)\times SU(2)$ gauge symmetry in the corresponding monopole background. Since this contains a product of two bosonic BPS letters, these carry gauge charges $(-1,-1)$ for ${\rm det}A$. An analogous statement is true for ${\rm det}\,B$ appearing in the $(1,1)$ flux sector. Here ${\rm det} A$ is an abbreviation for the operators $\epsilon_{ab}\epsilon_{\bar a\bar b} A^i_{a\bar a}A^j_{b \bar b}$ transforming as $\SUSU{3}{1}$ and similarly for ${\rm det}\,B$.}

In terms of $SU(4)$ representation theory, we have
\begin{equation}
    \begin{split}
        \big(YYYY\big)\big(YY\big) &:\, \overline{\mbf{10}}\times\overline{\mbf{35}} = \overline{\mbf{84}^{''}}+\overline{\mbf{140^{''}}}+\overline{\mbf{126}} ~, \\
        \big(YY\big)\big(YY\big)\big(YY\big) &:\, \overline{\mbf{10}}\times\overline{\mbf{10}}\times\overline{\mbf{10}} = \overline{\mbf{84}^{''}}+\overline{\mbf{10}}+\overline{\mbf{126}} ~.
    \end{split}
\end{equation}
Now the Hilbert series predicts a total of $434$ multi-trace chiral ring operators \cite{Cremonesi:2016nbo}. We can match this counting by treating the two $\overline{\mbf{126}}$ and $\overline{\mbf{10}}$ to be linked by trace relations. This way, we are left with operators in two $\overline{\mbf{84^{''}}}$, one $\overline{\mbf{140^{''}}}$, and a $\overline{\mbf{126}}$ representations, which exactly sums up to $434$ operators. Since we do not know the form of the trace relations, the operators are, in principle, allowed to be linear combinations of $(YYYY)(YY)$ and $(YY)(YY)(YY)$, where the precise linear combinations can only be fixed by the explicit action of $\mathcal Q$ on the dressed monopole operators:
\begin{equation}\label{eq:N2k2_YYYYYY}
    \overline{\mbf{140}^{''}} + \overline{\mbf{126}} + 2\times\overline{\mbf{84}^{''}} :\,\big(YYYY\big)\big(YY\big)+\big(YY\big)\big(YY\big)\big(YY\big)
\end{equation}
The full list of dressed monopole operators of this form is summarised in Table \ref{table_N2k2_orderx3}. The ones corresponding to the trivial flux sector have already been discussed in Table \ref{table2}

Let us now count fermionic single-trace operators that schematically take the form $(\Psi YYY)$. Based on $Y$ and $\Psi$ transforming in $\overline{\mbf 4}$ and $\mbf 4$ representations, respectively, we have that
\begin{equation}\label{Psirep}
    \big(\Psi YYY\big) :\, \mbf 4 \times \overline{\mbf 4} \times \overline{\mbf 4} \times \overline{\mbf 4} = \overline{\mbf{70}} + 2\times\mbf{64} + 3\times\overline{\mbf{10}} + \mbf{10} + 3\times\mbf 6 ~.
\end{equation}
We claim that the  BPS combinations of these operators is the $\overline{\mbf{70}}$ of $SU(4)$,
\begin{equation}\label{eq:N2k2_PsiYYY}
    \overline{\mbf{70}} :\, (\Psi YYY).
\end{equation}
We will give various arguments to explain this. We first use a representation theoretic argument that relies on supercharges that commute with $\mathcal Q$ to find the right number of these operators. We can organize the BPS cohomology in supermultiplets with respect to the superalgebra that commutes with the $\mathcal N=2$ subalgebra generated by $\mathcal Q$, as discussed in Appendix \ref{app:ABJM}. In particular, there are $6$ supercharges, $\mathcal Q^{\rm com}$, that commute with  $\mathcal Q$. These transform under the $SU(4)$ flavour symmetry in the representation $\mbf 6$. The  BPS single trace operators fill supermultiplets of the commutant subalgebra. Representation theory then predicts exactly $\overline{\mbf{70}}$ fermionic operators at order $x^3$. See Appendix \ref{app:single_grav} for more details. We can also reach the same conclusions by considering the explicit action of $\mathcal Q^{\rm com}$ on the single trace $(YYYY)$ at order $x^2$ which produces the operators $(\Psi YYY)$ at order $x^3$.

Using the $SU(4)$ representation theory, we have
\begin{equation}
    \mathcal Q^{\rm com} \big(YYYY\big) :\, \mbf 6\times\overline{\mbf{35}} = \overline{\mbf{140^{''}}} + \overline{\mbf{70}} ~.
\end{equation}
Comparing with \eqref{Psirep}, we see that the only common representation is the $\overline{\mbf{70}}$, which is therefore the only multiplet in the BPS cohomology of $\mathcal Q$. Last but not least, we can also check that this prediction is consistent with the counting of dressed monopole operators using Table \ref{table_N2k2_orderx3}.

Finally, we consider the fermionic multi-trace operators, which can be viewed as the product of $(\Psi Y)$ and $(YY)$ from the lower orders. As $SU(4)$ representations, we have that
\begin{equation}\label{eq:N2k2_Psi Y_YY}
    \mbf{15}\times\overline{\mbf{10}} = \overline{\mbf{70}} + \mbf{64} + \overline{\mbf{10}} + \mbf 6 :\, \big(\Psi Y\big)\big(YY\big)  ~.
\end{equation}
Comparing this with the superconformal index \eqref{eq:N2k2_index}, we conclude that all of these are independent, {\it i.e.}, there are no relations between them. On a  general ground, we expect relations between fermionic operators to appear at higher orders, since traces of four letters are not decomposable. The explicit counting of the dressed monopole operators is presented in Table \ref{table_N2k2_orderx3} in Appendix \ref{app:dressed}, where we also discuss some subtleties.

We claim that these are all multi-graviton BPS cohomologies up to order $x^3$. We provide a schematic counting summary in Table \ref{table_N2k2}.
\begin{table}[!h]
    \begin{center}
        \begin{tabular}{|c|c|c|c|c|} 
            \hline
            $x^{\mathcal D+J_3}$ & Operator & $(-1)^F$\,Total & $SU(4)_{a_i}$ rep. & Reference \\
            \hline\hline
            $x$ & $\big(YY\big)$ & $10$ & $\overline{\mbf{10}}$ & \eqref{eq:N2k2_YY} \\
            \hline\hline
            \multirow{3}{*}{$x^2$} & $\big(YYYY\big)$ & $35$ & $\overline{\mbf{35}}$ & \eqref{eq:N2k2_YYYY} \\
            & $\big(YY\big)\big(YY\big)$ & $55$ & $\overline{\mbf{35}}+\mbf{20'}$ & \eqref{eq:N2k2_YY_YY} \\
            & $\big(\Psi Y\big)$ & $-15$ & $-\mbf{15}$ & \eqref{eq:N2k2_PsiY}  \\
            \hline\hline
            \multirow{6}{*}{$x^3$} & $\big(YYYY\big)\big(YY\big)$ & \multirow{2}{*}{$434$} & \multirow{2}{*}{$\overline{\mbf{140^{''}}}+\overline{\mbf{126}}+2\times\overline{\mbf{84^{''}}}$} & \multirow{2}{*}{\eqref{eq:N2k2_YYYYYY}} \\
            & $\big(YY\big)\big(YY\big)\big(YY\big)$ & & & \\
            & $D\big(YY\big)$ & $10$ & $\overline{\mbf{10}}$ & \eqref{eq:N2k2_DYY} \\
            & $\big(YDY+\Psi\Psi\big)$ & $6$ & $\mbf 6$ & \eqref{eq:N2k2_DYY} \\
            & $\big(\Psi YYY\big)$ & $-70$ & $-\overline{\mbf{70}}$ & \eqref{eq:N2k2_PsiYYY} \\
            & $\big(\Psi Y\big)\big(YY\big)$ & $-150$ & $-\overline{\mbf{70}}-\mbf{64}-\overline{\mbf{10}}-\mbf 6$ & \eqref{eq:N2k2_Psi Y_YY} \\
            \hline
        \end{tabular}
    \end{center}
    \caption{Multi-graviton cohomologies in the strongly coupled $U(2)_k\times U(2)_{-k}$ ABJM theory at $k=2$.}\label{table_N2k2}
\end{table}
Thus, the multi-graviton index for the $U(2)_\times U(2)_{-2}$ ABJM theory can be expressed as
\begin{equation}\label{eq:N2k2_multigrav_index}
    \begin{split}
        \mathcal I_{N=2,k=2}^{\rm multi-grav}(x;a_i) = 1 &+ \chiSU{10} \,x + \Big( 2\chiSU{35} + \chi_{\mbf{20}^{'}}^{SU(4)} - \chi_{\mbf{15}}^{SU(4)} \Big)\,x^2 \\
        &+ \Big( 2\chiSU{84^{''}} + \chiSU{140^{''}} + \chiSU{126} - 2\chiSU{70} - \chi_{\mbf{64}}^{SU(4)}  \Big) \,x^3 \\[0.6mm]
        &+ \mathcal O(x^4) ~.
    \end{split}
\end{equation}

\subsection{Fortuitous cohomologies}

Comparing the multi-graviton index \eqref{eq:N2k2_multigrav_index} with the full superconformal index \eqref{eq:N2k2_index} up to order $x^3$ we find
\begin{equation}
    \mathcal I_{N=2,k=2}(x;a_i) - \mathcal I_{N=2,k=2}^{\rm multi-grav}(x;a_i) = - \chiSU{10}(a_i) \, x^3 + \mathcal O(x^4) ~.
\end{equation}
This is almost identical to the structure we observed in the $N=2$ ABJM theory at large $k$, wherein we constructed the fortuitous operators $\mathcal O_f$. Here we see that the representation $\overline{\mbf{10}}$ of $SU(4)$ at order $x^3$.

According to the branching rules, this representation decomposes into $SU(2)_u\times SU(2)_v\times U(1)_t$ as
\begin{equation}
    \overline{\mbf{10}} = \SUSU{3}{1}_{-2} + \SUSU{2}{2}_0 + \SUSU{1}{3}_2 ~.
\end{equation}
Here, the $\SUSU{2}{2}$ in the trivial flux sector are exactly the four large $k$ fortuitous operators $\mathcal O_f$. Recall that $\mathcal O_f$ effectively contained the product of the non-BPS current $\psi_{\rm non-BPS}$ with a meson in the $\SUSU{2}{2}$ representation. With this understanding, the 10 fortuitous operators can be schematically considered to be the product of the non-BPS fermionic current with the 10 $\big(YY\big)$ mesons combined with non-BPS operators of the form $(\Psi YYY)$ in the $\overline{\mbf{10}}$ of $SU(4)$. 

\section{Discussion and Outlook \label{sec:6}}

In this paper, we have studied the 1/12-BPS (or 1/16 for $k=1,2$) cohomologies for ABJM at $N=1,2$. Our goal was to identify the fortuitous states in this theory. At $N=1$, we found many fortuitous operators for $k>1$ essentially due to the extreme trace-relations. At $k=1$, the superconformal index is freely generated and this does not yield any fortuitous operators.  At $N=2$, we found fortuitous cohomologies both at large $k$ and at $k=2$. These cohomologies appear already at third non-trivial order in the superconformal index, which is much lower than in $\mathcal{N}=4$ SYM. As such, they admit a much simpler expression. We were also able to find fortuitous cohomologies at $k=2$, where monopole operators are needed. Leveraging the enhanced supersymmetry and associated R-symmetry, we found 10 fortuitous operators at third order. These 10 operators correspond to the $SU(4)$ R-symmetry completion of the 4 operators discovered at large $k$. The 6 new operators are thus monopole sectors operators. We conclude with some open questions.

\subsection*{Higher orders}

A natural next step is to continue our investigation but with higher quantum numbers, and other values of $k$ and $N$. It would be in particular interesting to study the theory at $k=1$. In this case, the theory corresponds to the dimensional reduction of $\mathcal{N}=4$ SYM and one would expect a lot of the properties to be similar in both theories. Note that at $k=1$, four operators saturate the unitarity bound and thus decouple as a free sector in the theory. It would be interesting to see how this affects the structure of fortuitous states.\footnote{It would be also interesting to approach the index for $k=1$ using the dual ADHM model. Recent results about the counting of states in this model have appeared in \cite{Hwang:2025hfs}.} In a different direction, working at $N=2$ and large $k$, one can study the structure of the cohomologies at higher quantum number. This will be presented in \cite{us2026}. Similar structure to those found in the BMN sector of $\mathcal{N}=4$ SYM seem to be present: one can act on core fortuitous operators with graviton operators \cite{Choi:2023znd,Choi:2023vdm}. The obvious interpretation of these operators is in terms of a graviton gas surrounding black holes. These would be of the grey galaxy type \cite{Kim:2023sig,Bajaj:2024utv,Choi:2025lck}, and solutions of this kind have recently been discussed in AdS$_4$ \cite{Jones:2025gno}.

\subsection*{Direct exploration of low $k$ cohomologies}

Another interesting direction to explore is to study the low-k cohomologies directly by the action of the supercharge. In this paper, to find fortuitous operators at $k=2$, we employed an indirect method. We used our knowledge of the large $k$ fortuitous operators and the enhanced supersymmetry (and associated R-symmetry) to $SU(4)$ complete the large $k$ spectrum. Up to third order in the superconformal index, this completion based on group theory matches the full index, showing strong indication that we have correctly identified the complete set of both monotone and fortuitous operators.

It would be preferable to work directly with the supercharge action, but at small $k$, this requires understanding how to act with the supercharge in the monopole sectors, which is currently an open problem (independently of fortuity). In particular we noticed some puzzles in the counting of dressed monopole operators at order $x^3$, as discussed in Appendix \ref{app:dressed}, which could be related to the action of the supercharges on monopoles. A more complete analysis could be possibly done by generalizing the methods of  \cite{Benna:2009xd,Dyer:2013fja,Pufu:2013eda}.

\subsection*{An entropy puzzle}

One of the results we have found is that the number of fortuitous operators at third non-trivial order jumps from four to ten as $k$ decreases from a large number to $k=2$. This makes sense at the intuitive level: as $k$ decreases, more and more monopole sectors can contribute to a fixed quantum number degeneracy. For graviton states, the $KK$ analysis tells us that this is indeed the expected behaviour: the spectrum of graviton operators reproduces the KK spectrum on AdS$_4 \times S^7/\mathbb{Z}_k$, and there are more operators the smaller $k$ is since the quotient group gets smaller. 

Surprisingly, for black hole states, one expects an opposite result. The Bekenstein-Hawking formula implies that we expect a degeneracy that scales as \cite{Aharony:2008ug}
\be
\log \rho  \sim N^{3/2} \sqrt{k} f(\Delta/N^{3/2}) \,,
\ee
where $f$ is an order one function for $\Delta \sim N^{3/2}$. It is a function that relates the area and mass of the appropriate extremal black holes in AdS$_4$. In the type IIA limit, we get
\be
\log \rho  \sim \frac{ N^{2}}{\sqrt{\lambda}}  \tilde{f}(\Delta/N^{2}) \,,
\ee
where again, $\tilde{f}$ is order one, and $\lambda=N/k$. In both cases, we see that the entropy \textit{decreases} as $k$ is decreased. Similar expressions were found for the topologically twisted index in \cite{Benini:2015eyy,Bobev:2022eus} and for the superconformal index in the Cardy limit in \cite{Choi:2019zpz,Choi:2019dfu,Bobev:2022wem}.

These results are surprising given both our explicit results for the cohomologies at $N=2$ and the structure of the superconformal index as a sum over monopole sectors. At the level of our results, one can of course argue that we are not in any asymptotic regime, neither in $N$ nor $\Delta$, so one should not imagine that the black hole entropy formula should apply. But the second point is much more general, and the expected scaling of the degeneracies mean that there must be gigantic cancellations between the bosonic and fermionic states sitting in various monopole sectors. 

One way to think about this, is in terms of a ``lifting" problem. One cannot call this an honest lifting problem, since there is no continuous coupling, but one can still imagine that as $k$ is decreased, more monopole operators descend to low scaling dimensions. As they do so, they can pair up with operators of the zero monopole sector (or more generally to lower monopole sectors) and form long representations. The particularity of ABJM is that $k$ labels both a coupling constant and a quantity that dictates which operator are present in the theory. If this intuition is correct, then the decrease of the entropy is due to the fact that as $k$ decreases, many more operators that were BPS at large $k$ have the opportunity to pair up and become long. It would be very interesting to see explicit examples of this, by studying the concrete supercharge action in the monopole sectors.

\section*{Acknowledgments}
It is a pleasure to thank Nikolay Bobev, Alba Grassi, Monica Guica,  Shota Komatsu, Ji Hoon Lee, Juan Maldacena, Kyriakos Papadodimas, Luigi Tizzano, and especially William Harding and Noppadol Mekareeya for useful discussions. AB, PS, AZ and RV are partially supported by the INFN, and PS and AZ by the MIUR-PRIN grant 2022NY2MXY (finanziato dall’Unione europea - Next Generation EU, Missione 4 Componente 1 CUP H53D23001080006).

\appendix

\section{Lagrangian description of ABJM} \label{app:ABJM}

In this appendix, we review a Lagrangian formulation of the $U(N)_k\times U(N)_{-k}$ ABJM theory \cite{Aharony:2008ug}. After the explicit action, we write down the explicit supersymmetry transformations of the basic fields and their charges under various symmetries. We express the action in a manifestly $SU(4)_R\equiv SO(6)_R$ notation. To this end, we define the bosonic and fermionic vectors
\begin{equation}\label{eq:notation_C_Upsilon}
    C_I = \left(\phi_1,\phi_2,\widetilde\phi_1^\dagger,\widetilde\phi_2^\dagger\right) \;,\; \Upsilon_\alpha^I = \left(-\lambda_2,\lambda_1,-\widetilde\lambda_2^\dagger,\widetilde\lambda_1^\dagger\right){}_\alpha ~,
\end{equation}
where both $C_I$ transforms in the fundamental of $SU(4)_R$, while $\Upsilon_\alpha^I$ transforms in the anti-fundamental representation of $SU(4)_R$. In addition to these, there are also two Chern--Simons gauge fields $A_\mu$ and $\hat A_\mu$. Furthermore, We use the conventions where $(\gamma^\mu)_{\alpha}^{\;\;\beta} = \{i\sigma_2, \sigma_1, \sigma_3\}$, such that these gamma matrices satisfy $\{\gamma^\mu, \gamma^\nu\} = 2\eta^{\mu\nu}$. Furthermore, we use $\epsilon_{\alpha\beta}$ and $\epsilon^{\alpha\beta}$ to lower and raise the spinorial indices such that $\epsilon_{\alpha\beta} \epsilon^{\beta\gamma} = \delta_\alpha^\gamma$ with $\epsilon^{12}=-\epsilon_{12}=1$. 

The action in terms of these fields can be expressed as the following \cite{Kwon:2009ar}
\begin{equation}
    \mathcal S = \int {\rm d}x^3 \big( \mathcal L_{\rm Kin} + \mathcal L_{\rm CS} + V_{\rm boson} + V_{\rm fermion} \big) ~.
\end{equation}
The kinetic term and the Chern--Simons can be expressed as
\begin{equation}
    \begin{split}
        \mathcal L_{\rm Kin} &= {\rm Tr}\left(-D_\mu C^{I\dagger} D^\mu C_I+ i\,\Upsilon^\dagger_I\gamma^\mu D_\mu\Upsilon^I\right) ~, \\
        \mathcal{L}_{\rm CS} &= \frac{k}{4\pi}\epsilon^{\mu\nu\rho} \,{\rm Tr} \left(A_\mu\partial_\nu A_\rho + \frac{2i}{3}A_\mu A_\nu A_\rho-\hat{A}_\mu\partial_\nu \hat{A}_\rho - \frac{2i}{3}\hat{A}_\mu \hat{A}_\nu \hat{A}_\rho \right) ~,
    \end{split}
\end{equation}
while the potentials can be expressed as:
\begin{equation}
    \begin{split}
        V_{\rm boson} &= \frac{4\pi^2}{k^2}{\rm Tr} \Big(C^{I\dagger} C_I C^{J\dagger} C_JC^{K\dagger} C_K +C_IC^{I\dagger} C_JC^{J\dagger} C_K C^{K\dagger} \\
        &\qquad\qquad\quad + 4\,C^{I\dagger} C_J C^{K\dagger} C_I C^{J\dagger} C_K - 6\,C_IC^{J\dagger} C_J C^{I\dagger} C_K C^{K\dagger} \Big) ~, \\
        V_{\rm fermion} &= \frac{2\pi i}{k} {\rm Tr} \big(C^{I\dagger} C_I \Upsilon^\dagger_J \Upsilon^J - C_I C^{I\dagger} \Upsilon^J \Upsilon^\dagger_J + 2\,C_I C^{J\dagger} \Upsilon^I \Upsilon^\dagger_J - 2\,C^{I\dagger} C_J \Upsilon^\dagger_I \Upsilon^J ~, \\
        &\qquad\qquad\quad + \epsilon_{IJKL}\,C^{I\dagger }\Upsilon^JC^{\dagger K}\Upsilon^L-\epsilon^{IJKL}C_I\Upsilon^\dagger_JC_K\Upsilon^\dagger_L \big) ~.
    \end{split}
\end{equation}
The action of the covariant derivatives on the scalar fields are defined as
\begin{equation}\label{eq:covder_action}
    \begin{split}
        D_\mu C_I &= \partial_\mu C_I + i (A_\mu C_I - C_I \hat{A}_\mu) ~,
        D_\mu C^I = \partial_\mu C^I + i (\hat{A}_\mu C^I - C^I A_\mu) ~, \\[1mm]
        D_\mu (C_IC^J) &= \partial_\mu(C_IC^J) + i \left[ A_\mu, C_IC^J \right] ~.
    \end{split}
\end{equation}
We also introduce the spinorial basis \cite{Dolan:2008vc}, which let us redefine the covariant derivative as:
\begin{equation}
    D_{\alpha\beta} = (\gamma^\mu)_{\alpha\beta}D_\mu \;,\; \quad F_{\alpha \beta} = \frac{i}{2} (\gamma^\mu \bar{\gamma}^\nu)_{\alpha \beta} F_{\mu \nu}, \quad \hat{F}_{\alpha\beta} ~.
\end{equation}
The $\mathcal N=6$ supersymmetry transformations can be expressed as the action of the $6$ supercharges on the fields. These supercharges live in the $\mbf 6$, {\it i.e.}, the antisymmetric representation of $SU(4)_R$. The explicit supersymmetry transformations can be expressed as \cite{Gaiotto:2008cg, Kwon:2009ar}:
\begin{equation}\label{eq:SUSY_transform}
    \begin{split}
        Q_{IJ} \,C_K &= \epsilon_{IJKL} \Upsilon^L \\
        Q_{IJ} \,C^K{}^\dagger &= \delta_I^K\Upsilon_J^\dagger - \delta_J^K\Upsilon_I^\dagger ~, \\
        (Q_{IJ})_\alpha (\Upsilon_{K})_\beta &= \epsilon_{IJKL} \,i\,\gamma^\mu_{\alpha\beta}D_\mu C^L{}^\dagger + \frac{2\pi i}{k} \epsilon_{\alpha\beta}\,\epsilon_{IJKL}\,\big(C^L{}^\dagger C_M C^M{}^\dagger - C^M{}^\dagger C_M C^L{}^\dagger \big) \\
        &\qquad\qquad\qquad\qquad\quad + \frac{4\pi i}{k}\epsilon_{\alpha\beta}\,\epsilon_{IJLM} C^L{}^\dagger C_K C^M{}^\dagger ~, \\
        (Q_{IJ})_\alpha (\Upsilon^K)_\beta &= \delta_I^K\left(i\gamma^\mu_{\alpha\beta}D_\mu C_J- \frac{2\pi i}{k}\epsilon_{\alpha\beta}(C_JC^{\dagger M}C_M-C_MC^{\dagger M}C_J)\right) \\
        &\qquad - \delta_J^K\left(i\gamma^\mu_{\alpha\beta}D_\mu C_I- \frac{2\pi i}{k}\epsilon_{\alpha\beta}(C_IC^{\dagger M}C_M-C_MC^{\dagger M}C_I)\right) \\
        &\qquad - \frac{4\pi i}{k}\epsilon_{\alpha\beta}\left(C_IC^{\dagger K}C_J - C_JC^{\dagger K}C_I\right) ~, \\
        Q_{IJ}A_\mu &= \frac{2\pi i}{k}\gamma_\mu \left(C_{[I}\Upsilon^\dagger_{J]} + \frac{1}{2}\epsilon_{IJKL} \Upsilon^K C^{\dagger L}\right) ~, \\
        Q_{IJ}\hat{A}_\mu &= \frac{2\pi i}{k}\gamma_\mu \bigg(\Upsilon_{[J}^\dagger C_{J]} + \frac{1}{2}\epsilon_{IJKL} C^{\dagger L} \Upsilon^{L}\bigg) ~.
    \end{split}
\end{equation}
Furthermore, we can define the transformations on the derivatives of the fields. Given the different nature of the fields, there are two possible definitions of the transformation under supersymmetry:
\begin{equation}
    \begin{split}
        Q_{IJ}( D_{\alpha\beta}\varphi) &= D_{\alpha\beta}(Q_{IJ}\varphi)+i(\gamma^\mu)_{\alpha\beta}\left((Q_{IJ}A_\mu)\varphi\mp\varphi(Q_{IJ}\hat{A}_\mu)\right)\\
        Q_{IJ}(D_{\alpha\beta}\tilde{\varphi}) &= D_{\alpha\beta}(Q_{IJ}\tilde{\varphi})+i(\gamma^\mu)_{\alpha\beta}\left((Q_{IJ}\hat{A}_\mu)\tilde{\varphi}\mp\tilde{\varphi}(Q_{IJ}A_\mu)\right)\\
    \end{split}
\end{equation}
where with $\varphi$ ($\tilde{\varphi}$) we denote bifundamental (anti-bifundamental) fields, and the choice of the sign depend on the action of the supercharge on a bosonic or fermionic field.

In the main text, we use two different bases for the Cartan of $SU(4)_R\cong SO(6)_R$, which are related as follows
\begin{equation}
    \begin{split}
         q_2 + q_3 &= R_4^4 - R_1^1 ~, \\
         q_1 - q_2 &= R_1^1 - R_2^2 ~, \\
         q_2 - q_3 &= R_2^2 - R_3^3 ~.
    \end{split}
\end{equation}
The charges of the various fields in the Lagrangian are listed in Table \ref{tab:ABJM_charges}.
\begin{table}[!h]
    \centering
    \begin{tabular}{|c|c|c|c|c|c|c|}
        \hline
        Field & $\mathcal D$ & $J_3$ & $q_1$ & $q_2$ & $q_3$ & $U(N)_k\times U(N)_{-k}$ \\
        \hline\hline
        $\phi_1$ & $\frac12$ & $0$ & $\frac12$ & $-\frac12$ & $-\frac12$ & $\SUSU{N}{\bar{N}}$\\
        $\phi_2$ & $\frac12$ & $0$ & $-\frac12$ & $\frac12$ & $-\frac12$ & $\SUSU{N}{\bar{N}}$ \\
        $\widetilde\phi_1^\dagger$ & $\frac12$ & $0$ & $-\frac12$ & $-\frac12$ & $\frac12$ & $\SUSU{N}{\bar{N}}$ \\
        $\widetilde\phi_2^\dagger$ & $\frac12$ & $0$ & $\frac12$ & $\frac12$ & $\frac12$ & $\SUSU{N}{\bar{N}}$ \\
        \hline
        $(\lambda_1)_\pm$ & $1$ & $\pm\frac12$ & $\frac12$ & $-\frac12$ & $\frac12$ & $\SUSU{N}{\bar{N}}$ \\
        $(\lambda_2)_\pm$ & $1$ & $\pm\frac12$ & $-\frac12$ & $\frac12$ & $\frac12$ & $\SUSU{N}{\bar{N}}$ \\
        $(\widetilde\lambda_1^\dagger)_\pm$ & $1$ & $\pm\frac12$ & $-\frac12$ & $-\frac12$ & $-\frac12$ & $\SUSU{N}{\bar{N}}$ \\
        $(\widetilde\lambda_2^\dagger)_\pm$ & $1$ & $\pm\frac12$  & $\frac12$ & $\frac12$ & $-\frac12$ & $\SUSU{N}{\bar{N}}$ \\
        \hline
        $A_\mu$ & 1 & $(1, 0, -1)$ & $0$ & $0$ & $0$ & $\SUSU{N^2}{1}$\\
        $\hat{A}_\mu$ & 1 & $(1, 0, -1)$ & $0$ & $0$ & $0$ & $\SUSU{1}{N^2}$\\
        $F_{\alpha\beta}$ & $2$ & $(1, 0, -1)$ & $0$ & $0$ & $0$ & $\SUSU{N^2}{1}$\\
        $\hat{F}_{\alpha\beta}$ & $2$ & $(1, 0, -1)$ & $0$ & $0$ & $0$ & $\SUSU{1}{N^2}$\\
        \hline
        $D_{\alpha\beta}$& $1$ & $(1, 0, -1)$ & $0$ & $0$ & $0$ & $\SUSU{1}{1}$\\
        \hline
        \hline
        $\mathcal{Q}$ & $1/2$  & $-1/2$ & $0$ & $0$ & $1$ & $\SUSU{1}{1}$\\ 
        $\mathcal{S}$ & $-1/2$  & $1/2$   & $0$ & $0$ & $-1$ & $\SUSU{1}{1}$\\
        \hline
    \end{tabular}
    \caption{The various fundamental fields of the Lagrangian and their associated charges/representations.}
    \label{tab:ABJM_charges}
\end{table}
From this table, we identify the letters as in \eqref{eq:BPS_words}, using which we can construct the BPS words, a gauge-invariant concatenation of letters that are closed under the one loop action of the supercharge.

\section{Multiplets using commutant supercharges} \label{app:single_grav}
 
In this work, we have focused on the $\frac{1}{12}$-BPS operators, that are annihilated by the chosen supercharge (and its conformal partner):
\begin{equation}
    \mathcal{Q} = Q_{34, -} \;,\; \mathcal{S} = S^{34, -} .
\end{equation}
The BPS letters are the ones that satisfy the BPS bound presented in \eqref{eq:BPS_cond}. At large $k$, the commutant $OSp(4|2)$ subalgebra can be defined by using \eqref{eq:algebra} \cite{Dolan:2008vc, Zwiebel:2009vb}. It follows that the supercharges that generate the (fermionic part of) this commutant subalgebra are:
\begin{equation}
    Q_{13, +}\;, \quad Q_{14, +}\;, \quad Q_{23, +}\;, \quad Q_{24, +} \quad \equiv Q_{IJ} ~.
    \label{eq:commutant}
\end{equation}

It is easy to determine the action of the commutant on the BPS letters by specialising the supersymmetry transformations in \eqref{eq:SUSY_transform}. For convenience, we choose to redefine the the supercharges into a $(\mbf 2,\mbf 2)$ representation:
\begin{equation}
    \widetilde Q_{i\,p}\equiv Q_{I=i,J=p+2} \;,\quad i=1,2\;,\;\; p=1,2 ~.
\end{equation}
The action of these supercharges on the BPS letters can then be expressed as\\

\begin{minipage}{0.45\textwidth}
    \centering
    Bifundamental fields
    \begin{equation}
        \begin{split}
            \widetilde Q_{pq}\,B_i &= \varepsilon_{pj}\varepsilon_{qi}\,\psi^{A}_j ~, \\
            \widetilde Q_{pq}\,\psi^{A}_i &= i\delta_{ip}\,DB_q ~.
        \end{split}
    \end{equation}
\end{minipage}
\hfill
\begin{minipage}{0.45\textwidth}
    \centering
    Anti-bifundamental fields
    \begin{equation}
        \begin{split}
            \widetilde Q_{pq}\,A_i &= \delta_{pi}\,\psi_q^{B} ~, \\
            \widetilde Q_{pq}\,\psi^{(B)}_i &= i\varepsilon_{pj}\varepsilon_{qi}\,DA_j ~.
        \end{split}
    \end{equation}
\end{minipage}\\

It can be seen from \eqref{eq:covder_action} that the action of the commutant supercharges commutes with that of the covariant derivative $D$.

For $k=1$ and $k=2$, the supersymmetry is enhanced to $\mathcal{N}=8$ and the corresponding superconformal algebra is $OSp(8|4)$ that corresponds to $SO(8)_R$ $R$-symmetry \cite{Aharony:2008ug, Kwon:2009ar}. We follow the same notation as in \cite{Bhattacharya:2008zy,Bhattacharya:2008bja}, where the eight supercharges for $k=1$ transform in the $(1,0,0,0)$ representation of $SO(8)_R$, and label the highest weight of the representation with the charges $(q_3,q_2,q_1,q_0)$ under the Cartan of $SO(8)_R$. The algebra for a generic $k$ is obtained by quotienting by the $\mathbb{Z}_k$ subgroup generated by the charge $q_0=4\pi/k$.

For $k>2$, the superconformal algebra reduces to $OSp(6|4)$, while for $k=2$ all the eight supercharges survive and hence we have $OSp(8|4)$. Since the supercharges $\mathcal{Q}$ and $\mathcal{S}$ used to define the index have charges $(\pm 1,0,0,0)$, the commutant algebra for $k=1$ and $k=2$ preserves an $SO(6)\cong SU(4)$ symmetry, and thus contain six supercharges transforming in the $\mbf{6}$ of $SU(4)$.\footnote{This $SU(4)$ should not be confused with the $SU(4)_R$ R-symmetry of the theory at generic $k$.} We do not provide the explicit supersymmetry transformations on the fields for $k=1,2$ as the two additional supercharges discussed above are constructed with dressed monopoles \cite{Klebanov:1998hh}.

In the main text, we found all the single-trace generators by an explicit computation, with the help of the $SU(2)\times SU(2)$ representation theory. In the following subsection, we will illustrate how the multiplet (under the corresponding commutant subalgebra) can be found by starting from the single-trace superconformal primaries and using the commutant algebra. We do this for the multiplet generated by $(AB)$ and match the result with the generators in Section \ref{sec:st_cohomologies}. We then discuss the general structure of the single-graviton multiplets in Subsection \ref{subsec:singlegrav}.

\subsection{Single-trace operators up to order $x^3$}

In this section, we demonstrate the construction of commutant subalgebra multiplet that can be constructed from the single trace $(AB)$ by the action of the commutant supercharges. The results of acting by a single supercharge are listed in Table \ref{tab:commutant*BA}. 
\begin{table}[!h]
	\begin{center}
		\begin{tabular}{c | c c c c} 
			& $Q_{13}$ & $Q_{14}$ & $Q_{23}$ & $Q_{24}$\\
			\hline
			$(B_1A_1)$ & $(B_1\psi_1^{(B)})$ & $(B_1\psi_2^{(B)}-\psi_2^{(A)}A_1)$ & $-$ & $(\psi_1^{(A)}A_1)$ \\
			$(B_1A_2)$ & $-$ & $(\psi_2^{(A)}A_2)$ & $(B_1\psi_1^{(B)})$ & $(B_1\psi_2^{(B)}+\psi_1^{(A)}A_2)$ \\
			$(B_2A_1)$ & $(B_2\psi_1^{(B)}+\psi_2^{(A)}A_1)$ & $(B_2\psi_2^{(B)})$ & $(-\psi_1^{(A)}A_1)$ & $-$ \\
			$(B_2A_2)$ & $(\psi_2^{(A)}A_2)$ & $-$ & $(B_2\psi_1^{(B)}-\psi_1^{(A)}A_2)$ & $(B_2\psi_2^{(B)})$ 
		\end{tabular}
		\caption{Single-trace operators obtained by applying all commutant supercharges to $(B_iA_j)$.}
		\label{tab:commutant*BA}
	\end{center}
\end{table}
Evidently, we na\"ively find eight disintct operators, but it is easy to see that there is linear relation between them. This linear dependent operator corresponds to $\psi_{\rm Non-BPS}$ in \eqref{eq:nonBPS_current}. This is as expected from the counting in the main text; the remaining seven exactly match the ones found using the $SU(2)\times SU(2)$ representation theory.

Applying another commutant supercharge to these seven operators yeild single-trace operators at order $x^3$. Since each operators appears twice in Table \ref{tab:commutant*BA}, we only need to apply two of the four charges. One can explicitly check that the following operators are obtained this way:
\begin{equation}
	iB_i{D}A_j+\psi_{j}^{(A)}\psi_{i}^{(B)} \;,\quad i{D}(B_i)A_j-\psi_{j}^{(A)}\psi_{i}^{(B)} \;,\quad {D}(B_iA_j) ~.
\end{equation}
Once again, these operators are linearly dependent, and by discarding one of them results in a perfect match with the operators in Section \ref{sec:st_cohomologies}. For single-trace primaries at order $x^2$ and higher, the procedure is identical and will not be reviewed here.

\subsection{Single graviton multiplets}
\label{subsec:singlegrav}

The BPS single-trace operators can be organised in supermultiplets of the commutant algebra. For simplicity, we start with $k=1$. The graviton multiplets are those that survive at large $N$ and can be identified from the KK spectrum of AdS$_4\times S^7$ given in Table 1 of \cite{Bhattacharya:2008zy}. The states that contribute to the index have $\delta=J+q_3- \mathcal{D}=0$.\footnote{$\Delta=0$ in Table 1 of \cite{Bhattacharya:2008zy}.} We can find these multiplet $S_n$ by decomposing the results in the aforementioned table, and we list these in Table \ref{TableSn}. The lowest state in each multiplet can be identified with the BPS operator $(YYY\ldots YY)$ containing $n$ letters $Y$ with complete symmetrisation. For $k=1$ all values of $n\ge 1$ appear once in the spectrum.
\begin{table}[!h]
    \begin{center}
        \begin{tabular}{|c|c|c|c|} 
            \hline
            $x^{\mathcal D+J_3}$  & $(-1)^F$ & $SO(6)$ rep. & $SU(4)$ rep.\\
            \hline
            $x^{\frac{n}{2}}$ & + & $(\frac{n}{2},\frac{n}{2},\frac{-n}{2})$ & $[0,0,n]$ \\
            \hline
            $x^{\frac{n}{2}+1}$ & - & $(\frac{n}{2},\frac{n}{2},\frac{2-n}{2})$ & $[1,0,n-1]$ \\
            \hline
            $x^{\frac{n}{2}+2}$ & + & $(\frac{n}{2},\frac{n-2}{2},\frac{2-n}{2})$  & $[0,1,n-2]$ \\
            \hline
            $x^{\frac{n}{2}+3}$ & - & $(\frac{n-2}{2},\frac{n-2}{2},\frac{2-n}{2})$ & $[0,0,n-2]$ \\
            \hline
        \end{tabular}
    \end{center}
    \caption{Graviton supermultiplets $S_n$ for $n\ge 1$. The $SO(6)$ highest weights are labeled by the charges $(q_2,q_1,q_0)$ and correspond to $SU(4)$ Dynkin labels $[q_0+q_1,q_2-q_1,q_1-q_0]$.}\label{TableSn}
\end{table}

We can obtain the $k=2$ multiplets by using the $\mathbb{Z}_2$ projection, where only those with even $n$ survive. Moreover, for $N=2$ and $k=2$, the multiplets $S_n$ with $n>4$ are decomposable and we are left with
\begin{equation}
    \begin{split}
        S_2 \;:\;\; &\overline{\mbf{10}}\, x + \mbf{15}\, x^2 + \mbf{6}\, x^3 + \mbf{1}\, x^4\, ~, \\
        S_4 \; : \;\; & \overline{\mbf{35}}\, x^2 +\overline{\mbf{70}}\, x^3 + \overline{\mbf{45}}\, x^4 + \overline{\mbf{10}}\, x^5 ~.
    \end{split}
\end{equation}
These can immediately be schematically identified with the operators 
\begin{equation}
    \begin{split}
        S_2 \;:\; &  (YY),(\Psi Y),(\Psi\Psi+YDY),(\Psi DY) ~, \\
        S_4\; : \; & (YYYY),(\Psi YYY),(\Psi\Psi YY+\ldots),(\Psi\Psi\Psi Y +\ldots) ~,
    \end{split}
\end{equation}
which is in agreement with the results in the main text and the list of single trace gravitons given in Table \ref{table_N2k2}.

For $k>2$, the $\mathbb Z_k$ projection we should only keep the states with $2\,q_0\equiv0({\rm mod}\,k)$ with considerable complications. In the large $k$ limit, we can simply set $q_0=0$ corresponding to keeping only the states with vanishing $U(1)_t$ charge in the decomposition $SU(4)\rightarrow SU(2)_u \times SU(2)_v\times U(1)_t$. For $N=2$ and $k=2$, we thus obtain
\begin{equation}
    \begin{split}
        S_2\;:\; &\SUSU{2}{2}\, x + \big(\SUSU{3}{1}+\SUSU{1}{3}+\SUSU{1}{1}\big)\, x^2 + \SUSU{2}{2}\, x^3 + \SUSU{1}{1}\, x^4 ~, \\
        S_4\;:\; & \SUSU{3}{3}\, x^2 +\big(\SUSU{4}{2}+\SUSU{2}{4}+\SUSU{2}{2}\big)\, x^3 + \big(\SUSU{3}{3}+\SUSU{3}{1}+\SUSU{1}{3}\big)\, x^4 \\
        &\hspace{7.5cm} + \SUSU{2}{2}\, x^5 ~,
    \end{split}
\end{equation}
which leads to following identifications:
\begin{equation}
    \begin{split}
        S_2\;:\; &  (AB)\,,\;(\psi^A A)\Big \vert_{\text{BPScurrents}}\;,\;\;(\psi^A\psi^B+ADB),(\psi^A DA)-(\psi^B DB)\Big \vert_{\text{singlet}}\\
        S_4\;:\; & (ABAB)\,,\;(\psi^A ABA)\,,\;(\psi^B\psi^A AB+\ldots)\,,\;(\psi^A\psi^A\psi^B A +\ldots)\
    \end{split}
\end{equation}
where we have been intentionally schematic and one should also consider terms with the swap of
$A$ and $B$ when applicable. The result is once again in agreement with Table \ref{table2}.

\section{Trace relations for $2\times 2$ matrices \label{app:tracerelations}}

In this appendix, we review some basic properties of $2\times 2$ matrices, in particular their trace relations. Using such relations we can  always reduce a trace containing $n$ matrices to a linear combinations of products of traces with at most three matrices. If the matrices commute we can further decompose the traces of three matrices down to products of at most two.  We refer the reader to \cite{Procesi1976,Razmyslov1974,DrenskyFormanek2004} for more details. We start by the first relation, which is valid for any $2\times 2$ matrix $M$. It reads
\be
    M^2=\Tr (M)M-\frac{1}{2}(\Tr^2 (M)-\Tr (M^2))\mathbb{1} \,.
\ee
This also generates relations for higher powers of the matrix, by iterating the $M^2$ relation. If we start multiplying different matrices together, then things get more complicated. But there are still relations if the matrices are arranged in particular ways, for example if we consider symmetric combinations of matrices. To see, this let us replace
\be
    M=M_1+M_2 \,.
\ee
On the right hand side, we will have
\be
M_1^2 + M_2^2 + M_1 M_2 + M_2 M_1 \,.
\ee
The terms $M_i^2$ can each be reduced using the identity for a single matrix, and hence we get an identity for the symmetric combination of any two $2\times 2$ matrix. We can then evaluate traces using these relations, and generate all kinds of relations.

In this paper, we explored up to cubic order in the superconformal index, which means we have at most three matrices. We do not assume that they commute, and therefore, in general,  they are no further decomposable. However we mostly find symmetric combinations, because antisymmetric combinations of letters are $\mathcal{Q}$-exact, due to the superpotential, see \eqref{eq:Superpot_relation}. Following the logic above, we can get relations for three matrices, and the one which was useful for us is 

\begin{equation} \label{tracerelation}
    \begin{split}
        {\rm Tr}\big(M_1M_2M_3\big) + {\rm Tr}\big(M_2M_1M_3\big) &= {\rm Tr}M_1\,{\rm Tr}\big(M_2M_3\big) + {\rm Tr}M_2\,{\rm Tr}\big(M_3M_1\big) \\
        &\quad\; + {\rm Tr}M_3\,{\rm Tr}\big(M_1M_2\big) - {\rm Tr}M_1 \, {\rm Tr}M_2 \, {\rm Tr}M_3 ~.
    \end{split}
\end{equation}

Setting $M_1=M_2$, we get the equation used in \eqref{eq:TraceRelation1}.

\section{Dressed monopole operators at order $x^3$}\label{app:dressed}

\begin{table}[!!!h]
    \begin{center}
        \begin{tabular}{|c|c|c|c|c|} 
            \hline
            $U(1)_t$ & Operator & $(-1)^F$\,Total & $SU(2)_u\times SU(2)_v$ \\
            
            \hline\hline
            \multirow{2}{*}{$3$} & $\big(T^{(3,0)}BBBBBB\big)^{(1)}$ & $7$ & $\SUSU{1}{7}$ \\
            & $\big(T^{(2,1)}B^{(1)}B^{(1)}B^{(1)}B^{(1)}B^{(2)}B^{(2)}\big)$ & $15$ & $\SUSU{1}{7}+\SUSU{1}{5}+\SUSU{1}{3}$ \\
            \hline\hline

            \multirow{5}{*}{$2$} & $\big(T^{(2,0)}ABBBBB\big)^{(1)}$ & $12$ & $\SUSU{2}{6}$ \\
            & $\big(T^{(2,0)}BBBB\big)^{(1)}\big(AB\big)^{(2)}$ & $20$ & $\SUSU{2}{6}+\SUSU{2}{4}$ \\
            & $\big(T^{(2,0)}BBB\psiA\big)^{(1)}$ & $-8$ & $-\SUSU{2}{4}$ \\
            & $(\det B\big)^2 (AB)$ & 24 & $\SUSU{2}{6}+\SUSU{2}{4} +\SUSU{2}{2}$ \\
            & $(\det B\big) (\epsilon\epsilon B\psi^A)$ & $-12$ & $-\SUSU{2}{4}-\SUSU{2}{2}$ \\
            \hline\hline

            \multirow{9}{*}{$1$} & $\big(T^{(1,0)}BBBABA\big)^{(1)}$ & $15$ & $\SUSU{3}{5}$ \\
            & $\big(T^{(1,0)}BBBA\big)^{(1)}\big(BA\big)^{(2)}$ & $32$ & $\SUSU{3}{5}+\SUSU{3}{3}+\SUSU{1}{5}+\SUSU{1}{3}$ \\
            & $\big(T^{(1,0)}BB\big)^{(1)}\big(BABA\big)^{(2)}$ & $27$ & $\SUSU{3}{5}+\SUSU{3}{3}+\SUSU{3}{1}$ \\
            & $D\big(T^{(1,0)}BB\big)^{(1)}$ & $3$ & $\SUSU{1}{3}$ \\
            & $\big(T^{(1,0)}B_1DB_2+T^{(1,0)}\psiA_1\psiA_2\big)^{(1)}$ & $1$ & $\SUSU{1}{1}$ \\
            & $\big(T^{(1,0)}B^{(1)}B^{(1)}\,{\rm BPSCurrent}^{(2)}\big)$ & $-21$ & $-\SUSU{3}{3}-\SUSU{1}{5}-2\times\SUSU{1}{3}-\SUSU{1}{1}$ \\
             & $\big(T^{(1,0)}B^{(1)}B^{(1)}\,{\rm BPSCurrent}^{(1)}\big)$ & $-17$ &  $-\SUSU{1}{5}-\SUSU{1}{3}-\SUSU{3}{3}$\\
            & $\big(T^{(1,0)}B\psi_A\big)^{(1)}\big(BA\big)^{(2)}$ & $-16$ & $-\SUSU{3}{3}-\SUSU{3}{1}-\SUSU{1}{3}-\SUSU{1}{1}$ \\
            & $\big(T^{(2,-1)}B^{(1)}B^{(1)}B^{(1)}B^{(1)}A^{(2)}A^{(2)}\big)$ & $15$ & $\SUSU{3}{5}$ \\

            \hline\hline
            \multirow{5}{*}{$0$} & $-$ & $60-48$ & $-$ \\
            & $\big(T^{(1,-1)}B^{(1)}B^{(1)}B^{(1)}A^{(1)}A^{(2)}A^{(2)}\big)$ & $24$ & $\SUSU{4}{4}+\SUSU{4}{2}$ \\
            & $\big(T^{(1,-1)}B^{(1)}B^{(1)}B^{(2)}A^{(2)}A^{(2)}A^{(2)}\big)$ & $24$ & $\SUSU{4}{4}+\SUSU{2}{4}$ \\
            & $\big(T^{(1,-1)}B^{(1)}B^{(1)}A^{(2)}\psi_B^{(2)}\big)$ & $-12$ & $-\SUSU{2}{4}-\SUSU{2}{2}$ \\
            & $\big(T^{(1,-1)}A^{(2)}A^{(2)}B^{(1)}\psi_A^{(1)}\big)$ & $-12$ & $-\SUSU{4}{2}-\SUSU{2}{2}$ \\
            \hline
            
        \end{tabular}
    \end{center}
    \caption{Multi-graviton cohomology representatives contributing to the $U(2)_2\times U(2)_{-2}$ ABJM theory at order $x^3$. For succinctness, we only report operators with positive $U(1)_t$ charge. Those with negative charge are simply obtained by swapping $A$ and $B$, and the two $SU(2)$ groups. The operators in the trivial monopole sector are not listed here, and we refer to Table \ref{table2}. In this table BPSCurrent$^{(i)}$ refers to the seven $\Psi Y$ operators of the abelian theories transforming in the $\SUSU{3}{1}+\SUSU{1}{3}+\SUSU{1}{1}$, thus with the exclusion of the accidental current $\psi_o$.}
    \label{table_N2k2_orderx3}
\end{table}

In this appendix, we present the dressed monopole operators contributing at order $x^3$. Using the logic used in \cite{Cremonesi:2013lqa}, we consider the residual massless fields in the monopole background. The method works very well for the Hilbert series of ABJM \cite{Cremonesi:2016nbo} and it has been used also for the superconformal index in \cite{Hayashi:2022ldo}. We expect this method to work well in enumerating the multi-graviton cohomologies as well, since we expect these to be captured by excitations along a classical background when we move in the moduli space.\footnote{Note however that there seems to be exceptions to this paradigm \cite{Chang:2025mqp,Kim:2025vup}.} Indeed, we observe see that we can reconstruct all the multi-graviton operators this way. It is less clear how to describe the fortuitous states that should appear in the monopole sector $\mbf{m}=\mbf{\widetilde{m}}=\pm(1,0)$ in terms of the residual abelianised theory, and discuss this below.

All the multi-graviton operators at order $x^3$ reported in Table \ref{table_N2k2} can be reconstructed from Table \ref{table_N2k2_orderx3}, which lists the dressed monopole operators. One can explicitly check that the $SU(2)\times SU(2)\times U(1)_t$ representations recombines into the $SU(4)$ representations
given in Table \ref{table_N2k2}.

It is important here to note that there are extra dressed monopole operators beyond the ones that we identify as multi-gravitons. In particular, we can construct the operators $T^{(-1,0)}A^{(1)}A^{(1)} \,\psi_o^{(i)}$ and $T^{(1,0)}B^{(1)}B^{(1)} \,\psi_o^{(i)}$ by using the singlet current associated to the accidental $U(1)_o$ symmetry of the abelian theories. These operators have the right form and the correct quantum numbers $\SUSU{3}{1}_{-1}+\SUSU{1}{3}_{1}$ to combine with the fortuitous states $\SUSU{2}{2}_{0}$ in the monopole sector to reconstruct the $\mbf{10}$ $SU(4)$-representation of fortuitous states. One linear combination can thus be identified with the fortuitous monopole operators.

However, a puzzle still remains: what happens to the other linear combination? This is not all, as there are a few other states that we can be constructed as dressing monopole operators, for example $T^{(-1,0)}A_1DA_2$, and its companion $T^{(1,0)}B_1DB_2$.\footnote{This state appears in the table in combination with a fermion bilinear but, in the reduced abelian theory, they are BPS by themselves.} These states are not counted by the index, there is not even the same number of bosons and fermions for such extra states, rendering it impossible for them recombine into $SU(4)$ representation. Therefore,  we believe that they do not contribute to the BPS cohomology. It would be interesting to understand the dynamical mechanism underlying their absence. In principle, such a physical argument may arise from dynamics of the  residual theory or a suitable modification of the supersymmetry transformations in the monopole backgrounds.

\bibliographystyle{ytphys}
\bibliography{ref}

\end{document}